\documentclass[11pt,a4paper]{article}
\pdfoutput=1

\usepackage{jcappub}
\usepackage{color}

\usepackage{wrapfig}
\usepackage{lscape}
\usepackage{rotating}

\usepackage{epstopdf}

{\count255=\time\divide\count255 by 60 \xdef\hourmin{\number\count255}
  \multiply\count255 by-60\advance\count255 by\time
  \xdef\hourmin{\hourmin:\ifnum\count255<10 0\fi\the\count255}}

\newcommand{\gsim}{\mbox{\raisebox{-1.ex}{$\stackrel
     {\textstyle>}{\textstyle\sim}$}}}

\newcommand{\impc}{\mathrm{Mpc}^{-1}}

\def\Neff{N_\mathrm{eff}}
\def\SigmaMnu{\sum m_\nu}

\begin{document}

\title{Testing for New Physics: Neutrinos and the Primordial Power Spectrum}

\author[a]{Nicolas Canac,}
\author[b]{Grigor Aslanyan,}
\author[a]{Kevork N. Abazajian,}
\author[c]{Richard Easther,}
\author[d]{and Layne C. Price}

\affiliation[a]{Department of Physics, University of California at Irvine, Irvine, CA 92697\vspace{4pt}}
\affiliation[b]{Berkeley Center for Cosmological Physics, University of California, Berkeley, CA 94720, USA\vspace{4pt}}
\affiliation[c]{Department of Physics, University of Auckland, Private Bag 92019, Auckland, New Zealand\vspace{4pt}}
\affiliation[d]{McWilliams Center for Cosmology, Department of Physics, Carnegie Mellon University, Pittsburgh, PA 15213, USA}

\emailAdd{ncanac@uci.edu}
\emailAdd{aslanyan@berkeley.edu}
\emailAdd{kevork@uci.edu}
\emailAdd{r.easther@auckland.ac.nz}
\emailAdd{laynep@andrew.cmu.edu}

\date{\today}

\abstract{We test the sensitivity of neutrino parameter constraints
  from combinations of CMB and LSS data sets to the assumed form of
  the primordial power spectrum (PPS) using Bayesian model selection.
  Significantly, none of the tested combinations, including recent
  high-precision local measurements of $\mathrm{H}_0$ and cluster
  abundances, indicate a signal for massive neutrinos or extra
  relativistic degrees of freedom.  For PPS models with a large, but
  fixed number of degrees of freedom, neutrino parameter constraints
  do not change significantly if the location of any features in the
  PPS are allowed to vary, although neutrino constraints are more
  sensitive to PPS features if they are known \emph{a priori} to exist
  at fixed intervals in $\log k$.  Although there is no support for a
  non-standard neutrino sector from constraints on both neutrino mass
  and relativistic energy density, we see surprisingly strong evidence
  for features in the PPS when it is constrained with data from
  \emph{Planck} 2015, SZ cluster counts, and recent high-precision
  local measurements of $\mathrm{H}_0$.  Conversely combining
  \emph{Planck} with matter power spectrum and BAO
  measurements yields a much weaker constraint.  Given that this
  result is sensitive to the choice of data this tension between SZ
  cluster counts, \emph{Planck} and $\mathrm{H}_0$ measurements is
  likely an indication of unmodeled systematic bias that mimics PPS
  features, rather than new physics in the PPS or neutrino
  sector. 
}
\arxivnumber{1606.03057}

\maketitle

\section{Introduction}

Cosmological observations have emerged as a stringent constraint on
the total mass of neutrinos. The total neutrino mass makes a subtle
contribution to cosmic microwave background (CMB) anisotropies and has
a more substantial impact on the late time matter power spectrum
measured via clustering observations.  The CMB is sensitive to
neutrino mass and effective neutrino number via the alteration of
matter-radiation equality leading up to the decoupling of the CMB, and
the alteration of the evolution of the neutrino anisotropic
stress-energy tensor. Extra relativistic energy density modifies the
acoustic peak scale to the photon-damping scale, making the CMB a
sensitive measure of relativistic energy density above the photon
density, often parameterized as $\Neff \equiv (\rho_{\mathrm{rad}} -
\rho_\gamma)/\rho_\nu$, where $\rho_{\mathrm{rad}}$, $\rho_\gamma$,
and $\rho_\nu$ are the total energy density in relativistic species,
photons, and active neutrinos, respectively.

Cosmological large scale structure (LSS) is  sensitive to the
presence of massive neutrinos and effective neutrino number
$\Neff$. Baryon acoustic oscillations (BAO) are affected by the change
of matter-radiation equality and commensurate change in expansion
history.  Measurements of LSS clustering, such as the power
spectrum of galaxies, are a sensitive probe of neutrino properties as
clustering is suppressed below the neutrino free streaming scale
via a combination of the relativistic behavior of neutrinos at
early times and their free-streaming suppression of late time growth.
The sensitivity of galaxy clustering was highlighted some time ago
(\emph{e.g.},~\cite{Hu:1997mj}), and future probes can achieve very
high precision (For a recent review, see
Ref.~\cite{AbazajianAnnRev:2016}).

When combined with LSS observables, the complementary role of the CMB
removes degeneracies with other cosmological parameters allowing
high-precision determinations of their values. For example, the scalar
perturbation amplitude $A_s$, tilt $n_s$ and matter density $\Omega_m$
are each, to different degrees, degenerate with $\SigmaMnu$, but are
determined to percent-level precision from Planck's 2015 CMB
analysis~\cite{Ade:2015xua}.  In single-parameter extensions of
$\Lambda$CDM in which the sum of neutrino masses is a free parameter,
the resulting constraints are an order of magnitude tighter than
current kinematic laboratory constraints: $\sum m_{\nu} < 0.23\rm\ eV$
(95\%) from Planck 2015 (TT, lowP, lensing, BAO, JLA,
$\mathrm{H}_0$)~\cite{Ade:2015xua}, versus $\sum m_{\nu} \lesssim
6\rm\ eV$ from $^3$H $\beta$ decay plus
oscillations~\cite{Agashe:2014kda}.  The latest Planck data (TT, TE,
EE, SimLow, lensing)~\cite{Aghanim:2016yuo} yields $\sum m_\nu <
0.14\ \mathrm{eV} (95\%)$, but the HFI likelihood codes are not yet
public.  Here the sum of neutrino masses is defined as the sum of the
individual mass eigenstates $\Sigma m_\nu \equiv m_1+m_2+m_3$ and does
not depend on their hierarchical ordering beyond the mass degeneracy
scale where $m_1\approx m_2\approx m_3$.

In the past several years, there have been indications of tension
between local probes of cosmological structure and expansion and
values obtained from analyses of CMB data. The tension comes in two
regards: first, the CMB-parameter inferred local Hubble expansion rate
at present, $\mathrm{H}_0$, is lower than precise local measures; and
second, the amplitude of scalar perturbations inferred from the CMB is
higher than that from more local measures. The combination of the CMB
with these local probes are often referred to as tension data sets and
we adopt that here.  These tension data sets include analyses of
combinations of CMB and $\mathrm{H}_0$ data with SPT SZ clusters by
Hou et al.~\cite{Hou:2012xq}, and combinations  of CMB data with cosmic shear lensing data
from CFHTLenS also indicated a nonzero neutrino mass
\cite{Battye:2013xqa}, as well as data from the Baryon Oscillation
Spectroscopic Survey (BOSS) constant mass (CMASS) luminous red galaxy
sample \cite{Beutler:2014yhv}. With such combinations of data sets
between local and high redshift cosmology measures, there were a
number of combinations that indicated degenerate neutrino masses or
extra relativistic energy density could relieve this tension
\cite{Wyman:2013lza,Verde:2013wza,Battye:2014qga,Giusarma:2014zza,Roncarelli:2014jla,Raveri:2015maa}. Whether
the evidence for strong tension and new neutrino physics from the low
redshift measures is definitive has been called into question by
several papers \cite{Verde:2013cqa,Leistedt:2014sia,Ade:2015xua}. For
example, weak lensing systematics has been shown to alleviate tension
in that data set \cite{MacCrann:2014wfa}. Generally, the low redshift
data indicates a lower amplitude of fluctuations on relatively small
scales, parameterized as $\sigma_8$, the rms over-density of
fluctuations smoothed with a spherical window function of $8
h^{-1}\ \mathrm{Mpc}$.  Although there exist several different cluster
abundance samples that indicate tension with the CMB
\cite{Saro:2013fsr,Ade:2013lmv,Vikhlinin:2008ym,Rozo:2009jj}, we
employ, as a representative measure, constraints on $\sigma_8$ vs
$\Omega_m$ from Planck SZ Clusters, as described below
\cite{Ade:2013lmv,Battye:2014qga}.  Additional tension is indicated in
recent high-precision measures of the local expansion rate
$\mathrm{H}_0$, which has been proposed to potentially indicate extra
relativistic energy density ($\Neff > 3$)
\cite{Riess:2016jrr}. Importantly, the tension between low and high
redshift perturbation amplitude is at least partially alleviated in
new polarization measures of reionization that reduce the inferred
scalar amplitude \cite{Aghanim:2016yuo}.  In summary, a number of tension
data sets have suggested non-standard relativistic energy or
non-trivial neutrino mass, but the results are inconclusive, nor are
necessarily mutually consistent. Such tension could very well indicate
new physics and is why we include it in our investigation.

Cosmological neutrino mass measurements are approaching the sensitivity
needed to detect the minimal value of $\SigmaMnu \approx\ 60\rm\ meV$
derived from the mass-splittings in the neutrino sector inferred from
neutrino oscillations. However, reliably achieving this sensitivity
requires a careful analysis of the assumptions and model dependencies
underlying cosmological constraints. Several cosmological model
dependencies are discussed in, \emph{e.g.}
Ref.~\cite{Abazajian:2011dt}. In particular, the effects of massive
neutrinos on large scale structure can be degenerate with deviations
from smoothness in the primordial power spectrum (PPS).
  In particular, given a fine-tuned PPS, CMB
  data can mimic a zero neutrino mass universe even if the neutrino
  density is large \cite{Kinney:2001js}.  It is therefore important to
  test the dependence of neutrino parameter constraints on the assumed
  shape of the PPS.  The consequences of a non-trivial PPS for
  neutrino constraints were previously analyseds in
  Refs.~\cite{dePutter:2014hza,Gariazzo:2014dla,DiValentino:2016ikp}.
  Those analyses consider a PPS in which features and/or
  discontinuities were located at specified wavenumbers. In this
  paper, we consider more general scenarios in which the features
  locations are constrained by the data, employing the Bayesian model
  selection methods of Ref~\cite{Aslanyan:2014mqa,Abazajian:2014tqa}
  to determine whether the improved fit justified the increased
  complexity of the model.

Separately, we go beyond previous work by exploring tension between SZ
cluster abundances and recent high-precision measures of
$\mathrm{H}_0$ in the ``local'' universe.  We combine the 2015 Planck
results \cite{Ade:2015xua}, large-volume galaxy survey LSS data from
the clustering of Luminous Red Galaxies from the Sloan Digital Sky
Survey \cite{Reid:2009xm}, and the WiggleZ Dark Energy Survey
\cite{Parkinson:2012vd}. We also use the most recent measures of the
Baryon Acoustic Oscillation (BAO) scale from the six-degree-Field
Galaxy Survey (6dFGS) \cite{Beutler:2011hx}, from the SDSS Main Galaxy
Sample (SDSS-MGS) \cite{Ross:2014qpa}, and from the Baryon Oscillation
Spectroscopic Survey data release 11 (BOSS DR11), from both the LOWZ
and CMASS samples \cite{Anderson:2013zyy}, which provide robust
complementary constraints on the cosmological parameters, including
neutrinos. In order to claim strong evidence for new physics from the
combined analysis of these tension data sets, it is necessary that
different probes of the same cosmological signals, \emph{e.g.},
expansion history or growth rate, provide comparable statistical
evidences for the same extensions to $\Lambda$CDM.  In contrast, if
different models are favored with wildly variable confidences, then
this provides some evidence for unmodeled bias in the probes
themselves.  In light of this, here we aim to test for the evidence of
novel signals in the neutrino sector or the primordial spectrum.

\section{Method}

In this section we begin by describing the cosmological models we analyze, along with 
the ``knot-spline'' algorithm used to reconstruct the PPS. We also
give our priors for the parameters in our models and discuss
the use of Bayesian evidence and posterior probabilities in
model selection and parameter estimation.

The likelihood calculations and PPS
reconstruction are performed using the publicly available code
\textsc{Cosmo++} \cite{Aslanyan:2013opa}. The CMB power spectra and
matter power spectrum are calculated using the \textsc{CLASS} package
\cite{Lesgourgues:2011re,Blas:2011rf}. The parameter space sampling
and Bayesian evidence calculation is implemented with the publicly
available multimodal nested sampling code \textsc{MultiNest}
\cite{Feroz:2007kg,Feroz:2008xx,Feroz:2013hea}. Finally, the resulting
chains are analyzed and plotted using the \textsc{GetDist} Python
package.

\subsection{Non-power-law primordial power spectrum}

Our goal is to examine how constraints on neutrino parameters change
when we relax the assumption that the PPS has a simple
power-law form.  In particular, we want to test the sensitivity of
neutrino parameters to the shape of
the PPS.  Our approach to reconstructing the PPS is a
variation of the ``knot-spline'' method  described in
\cite{Vazquez:2012ux,Aslanyan:2014mqa,Abazajian:2014tqa}.  Similar
methods implementing a free-form primordial spectrum are given  in
Refs.~\cite{Sealfon:2005em,Bridges:2005br,Bridges:2006zm,Verde:2008zza,Bridges:2008ta,Peiris:2009wp,Bird:2010mp,Vazquez:2011xa,dePutter:2014hza,Ade:2015lrj,DiValentino:2016ikp,Ravenni:2016vjd}.

The algorithm is 
summarized as follows:
\begin{enumerate}
\item Fix $k_\mathrm{min} = 10^{-6} \, \impc$ and $k_\mathrm{max}=10.0 \, 
\impc$, but allow their amplitudes $A_\mathrm{min}$ and $A_\mathrm{max}$ to 
vary.

\item Add $n$ knots with uniform priors on $\log k$, in the range  $\log 
k_\mathrm{min} < \log k_i < \log k_\mathrm{max}$ and a uniform prior on $A_i$, in 
the range $-2 < A_i < 4$, where $A_i \equiv \log(10^{10}\Delta^2_\zeta(k_i))$, 
where $\Delta^2_\zeta$ is the dimensionless PPS of curvature perturbations and
$i=1,2,...,n$. The knots are ordered so that $k_{i-1} \le 
k_i$, and the number of knots $n$ is varied between $1$ and $5$.

\item Interpolate between the endpoints and the $n$ ordered knots using a 
linear spline. The interpolation is performed in logarithmic space for both $k$ 
and $\Delta^2_\zeta$.

\end{enumerate}

\begin{figure}
\centering

\includegraphics[width=0.48\textwidth]{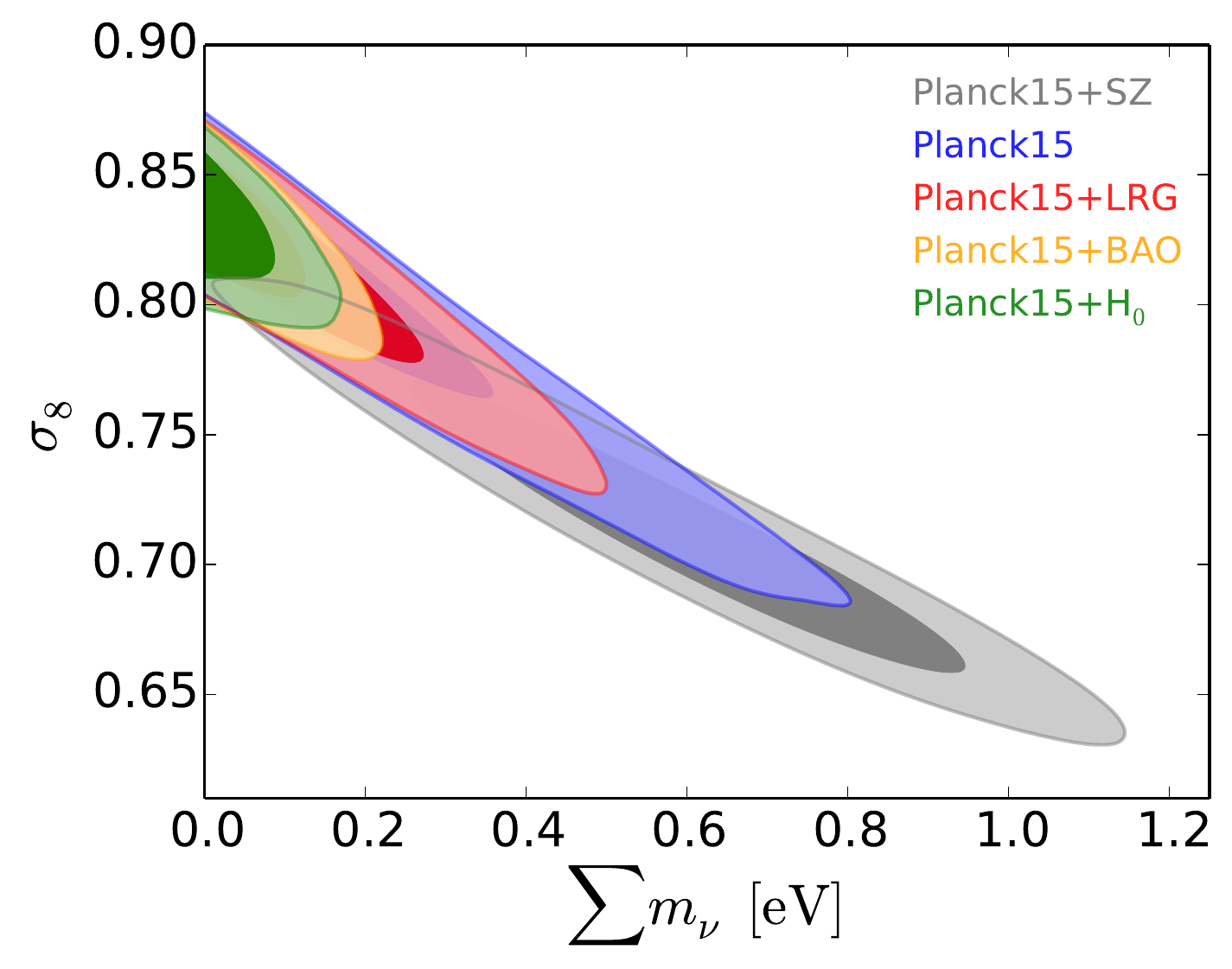}
\includegraphics[width=0.48\textwidth]{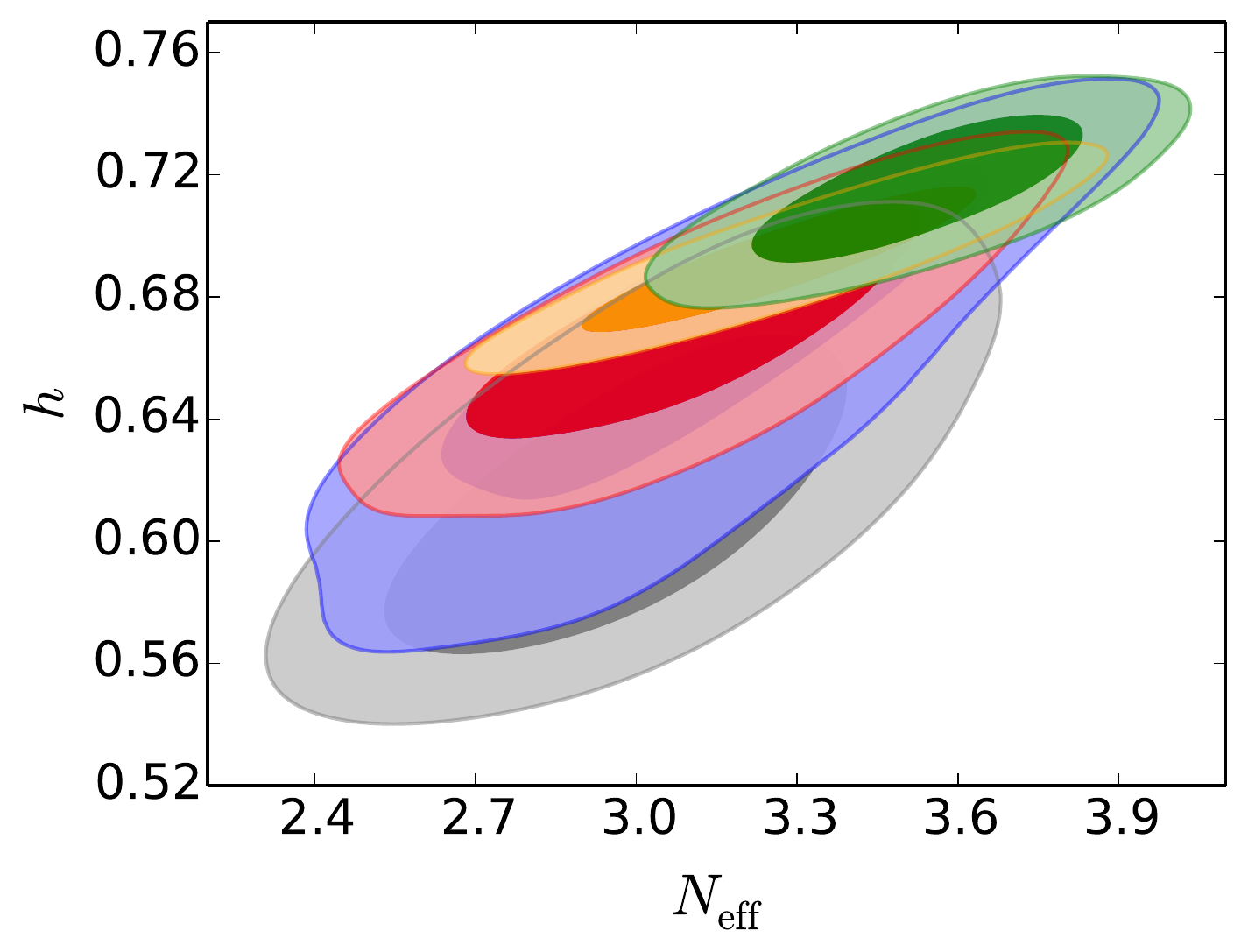}
\caption{\label{sigma8_v_summnu_0knots_fig} \emph{(Left)} The two-dimensional 
posterior distribution showing the $68\%$ and $95\%$ CI allowed regions in
the $\sigma_8-\SigmaMnu$ plane for 0 knots and $\SigmaMnu$ free for various 
combinations of data. \emph{(Right)} The same but for $\Neff$ in the 
$h-\Neff$ plane.}
\end{figure}

We adopt uniform priors for the locations of knots in $\log k$ and
their amplitudes.  Knots with arbitrary locations in $k$-space are
able to capture both local, step-like feature, more gradual changes
like a large-scale exponential suppression, or a spectrum with a sharp
cutoff.  With no knots, the PPS is specified by two parameters, the
amplitudes of the endpoints $k_\mathrm{min}$ and $k_\mathrm{max}$. In
this case, the PPS is equivalent to the standard power-law PPS in
$\Lambda$CDM, providing for an easy comparison between the two
models. Each additional knot yields two degrees of freedom
corresponding to the location of the knot $k_i$ and its amplitude
$A_i$. In total, $2n+2$ free parameters specify the PPS model, where
$n$ is the number of knots. Allowing the knot location to vary
provides some basic protection against the look-elsewhere effect or
multiple comparisons problem, since the knot is free to move over the
global range of $k$, in contrast to reconstructions in which the knot
locations are fixed. The broad ranges from which $k_{\mathrm{knot}}$
and $A_{\mathrm{knot}}$ are drawn allow for possible features at any
measurable scale.

\subsection{Model, priors, and Bayesian evidence}

We choose uniform priors for all cosmological variables,
including the usual parameters $\Omega_bh^2$, $\Omega_ch^2$, $h$,
and $\tau$, the sum of the neutrino masses $\SigmaMnu$
and the effective number of relativistic degrees of freedom
$N_\mathrm{eff}$. As explained above, the knot-spline specification of
the PPS has two parameters associated with each knot, $\log_{10}
k_\mathrm{knot}$ and $\log (10^{10} \Delta_\mathrm{knot}^2)$, and two
additional parameters describing the amplitudes of the fixed
endpoints. The ranges for these priors are shown in
Table~\ref{priors_table}.

\begin{figure}
\centering
\includegraphics[width=15cm]{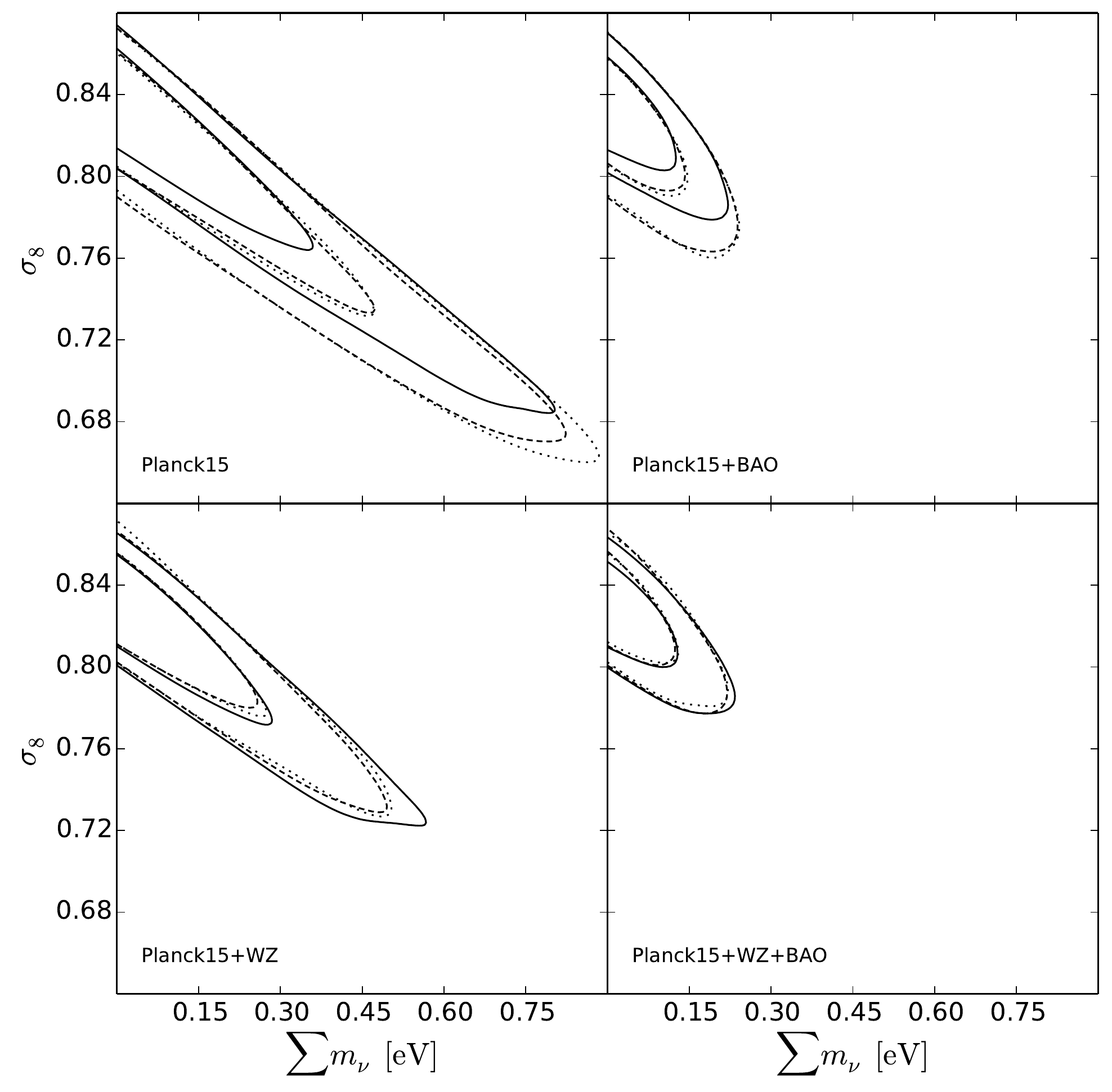}
\caption{\label{sigma8_v_summnu_comparison_fig} Two-dimensional posterior 
distributions in the $\sigma_8-\SigmaMnu$ plane for 0 knots (solid line), 1 knot (dashed 
line), and 2 knots (dotted line) for various combinations of data sets. Models with more than 2 knots do not differ significantly from the $n=2$ case and are not displayed.
}
\end{figure}

\begin{figure}
\centering
\includegraphics[width=15cm]{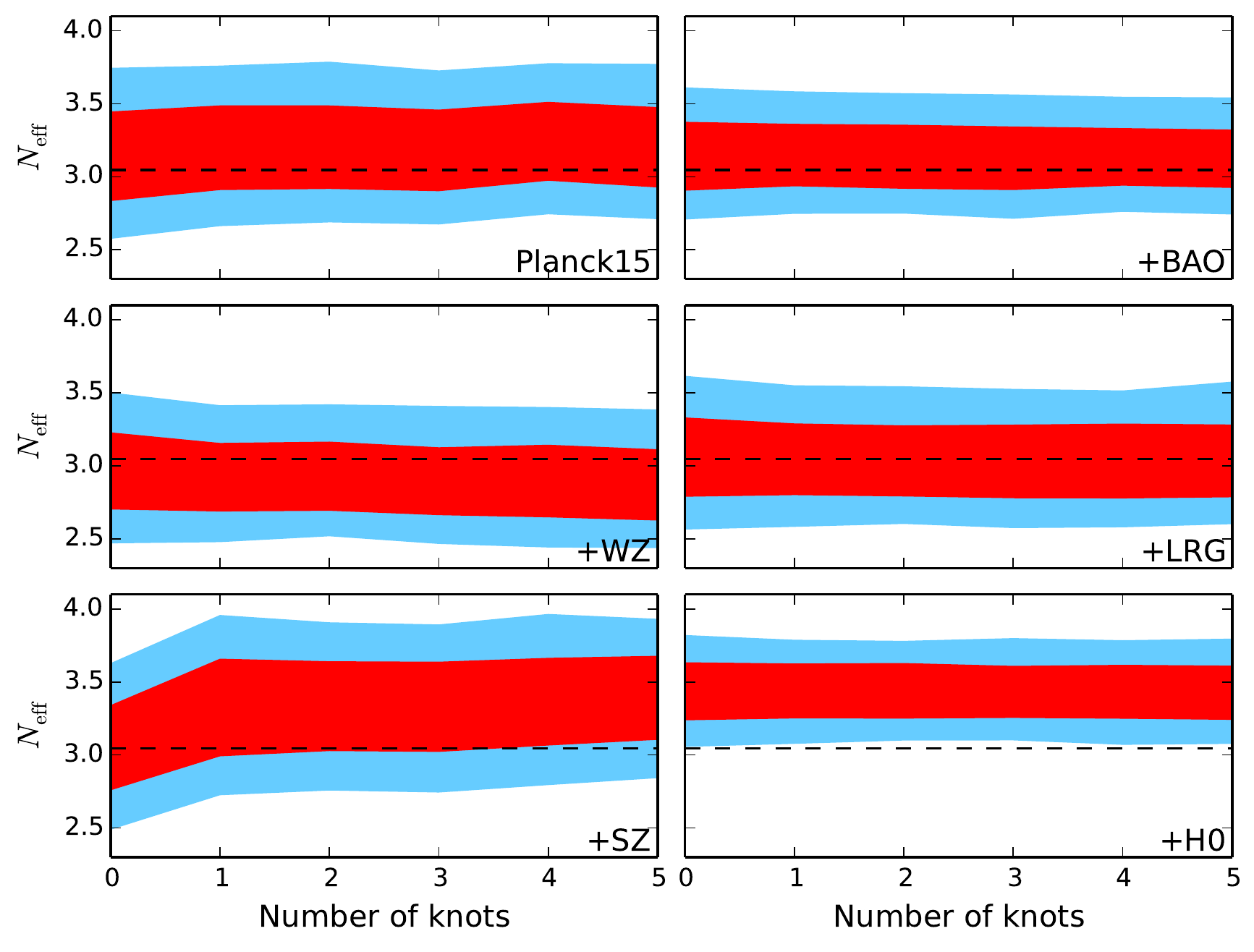}
\caption{\label{neff_lims_fig} $68\%$ and $95\%$ CI constraints on 
$N_\mathrm{eff}$ for models with $N_\mathrm{eff}$ allowed to vary and PPS 
reconstruction with knot location free. The data sets used are indicated in 
each panel (Planck15 is implicitly included in each panel).}
\end{figure}

To evaluate the statistical significance of a model $\mathcal M$, we use the 
posterior probability $P(\mathcal M \, | \, \mathrm{Data})$. For two models 
$\mathcal M_i$ and $\mathcal M_j$ with the same prior probability, the evidence 
ratio or Bayes factor is given by
\begin{equation}
  \frac{Z_i(D)}{Z_j(D)} = \frac{P(\mathrm{Data}\, | \, \mathcal M_i)}{P(\mathrm{Data}\, | \, 
  \mathcal M_j)} \, ,
  \label{eqn:bayes_factor}
\end{equation}
where the Bayesian evidence or marginalized likelihood is
\begin{equation}
  Z_i(D) \equiv P(\mathrm{Data} \, | \,\mathcal M_i) = \int P(\theta \, | \, 
  \mathcal M_i ) \, \mathcal L(\mathrm{Data} \, | \, \theta) d\theta
  \label{eqn:evidence}
\end{equation}
for the model parameters $\theta$. Here, $\mathcal L(\mathrm{Data} \,
| \, \theta)$ is the data-likelihood and $P(\theta \, | \, \mathcal
M_i )$ is the parameter prior probability. When the prior probabilities
for the models are equal (as is common convention) the
Bayes' factor directly measures the posterior model odds. We allow for a wide range of values when specifying potential
features in the PPS at fixed number of degrees of freedom in our
parameterization. In general, the integral in
equation~\eqref{eqn:evidence} is numerically challenging but can be
computed using multimodal nested sampling.

It is more convenient to use the logarithm of the Bayes factor:
\begin{equation}
  \Delta\ln Z\equiv\ln\frac{Z_i(D)}{Z_j(D)}\,.
\end{equation}

Evidence ratios can be interpreted qualitatively using the Jeffreys'
scale \cite{Jeffreys:1961} or a more conservative ``cosmology scale''
\cite{hobson2010bayesian}, summarized in
Table~\ref{jeffreys_table}. Bayesian model selection  yields a global estimate of how well a model fits the data  by integrating over the entirety of parameter
space and the Bayes factor  penalizes those complex models with many
free parameters
that yield a high value for the likelihood only within some small
sub-region of the overall parameter space.  A model with more
parameters must thus yield a significant improvement in likelihood over a
sufficiently large volume of parameter space in order to yield
posterior odds that support the more complex model.

\begin{figure}
\centering
\includegraphics[width=0.24\textwidth]{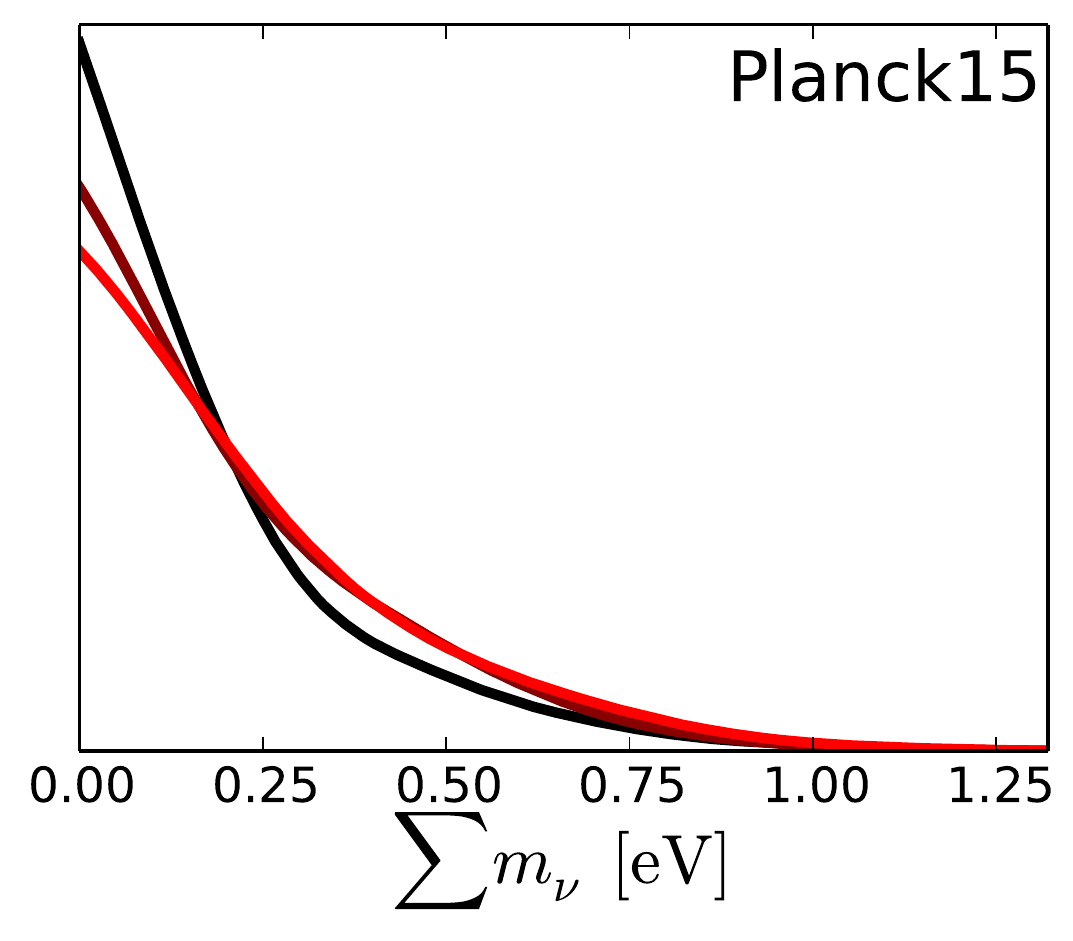}
\includegraphics[width=0.24\textwidth]{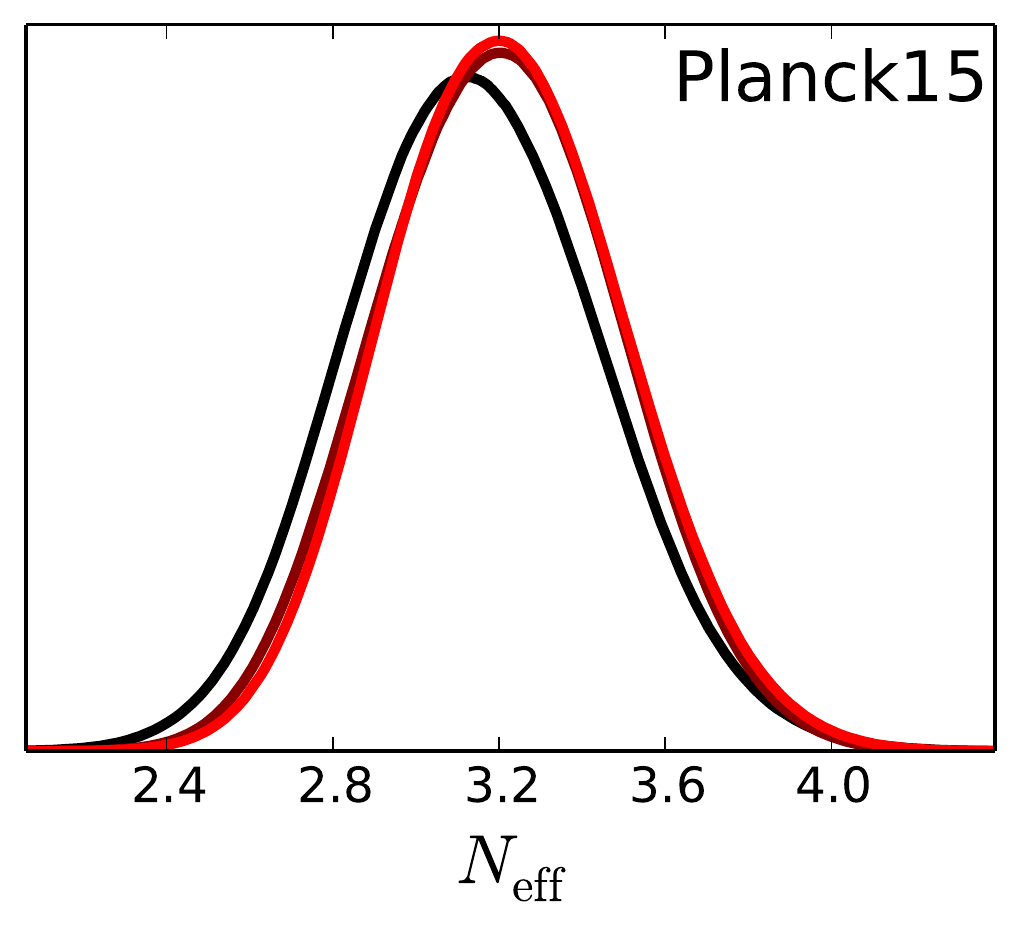}
\includegraphics[width=0.24\textwidth]{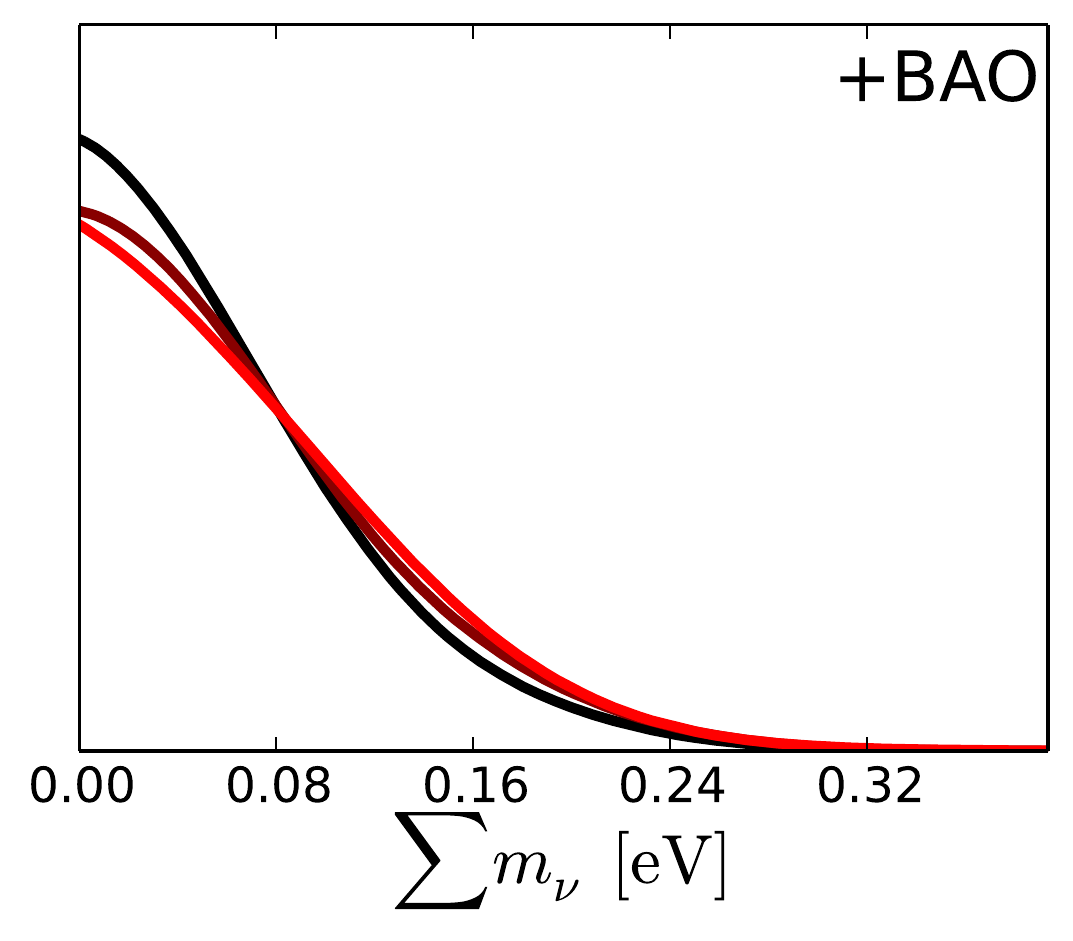}
\includegraphics[width=0.24\textwidth]{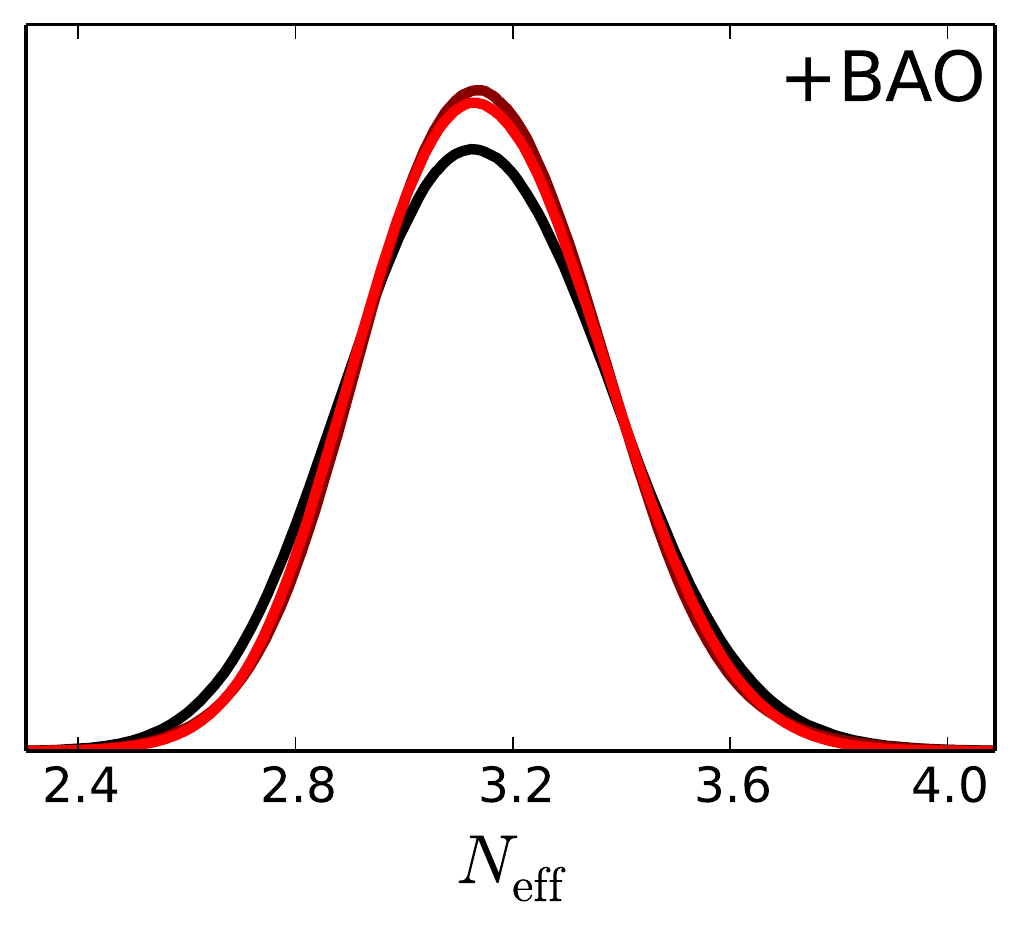}
\includegraphics[width=0.24\textwidth]{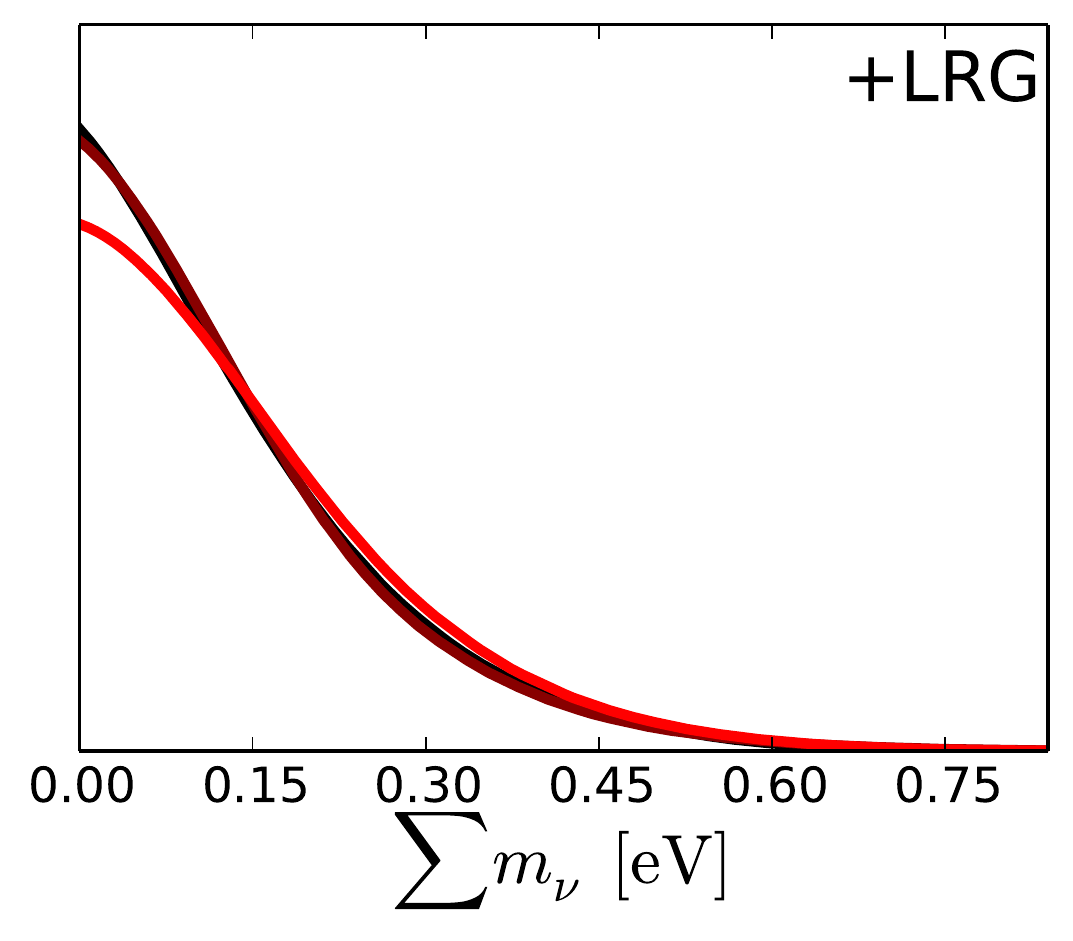}
\includegraphics[width=0.24\textwidth]{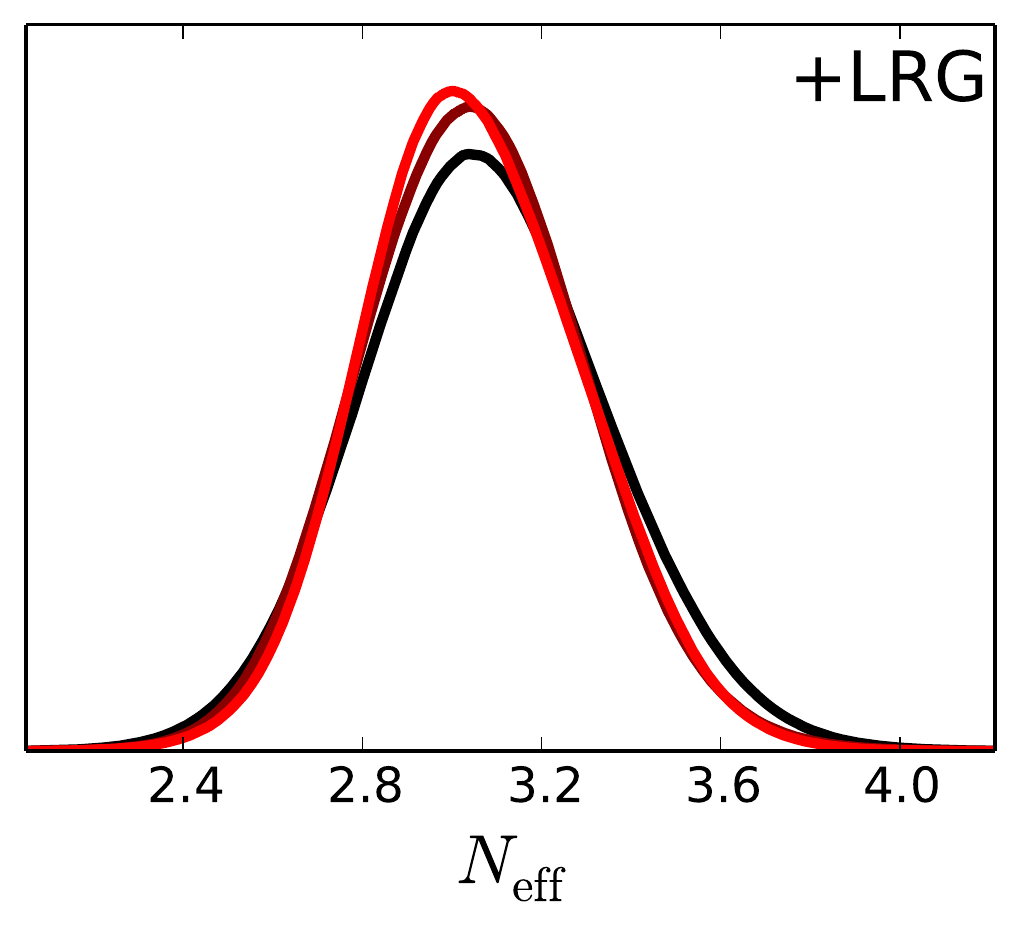}
\includegraphics[width=0.24\textwidth]{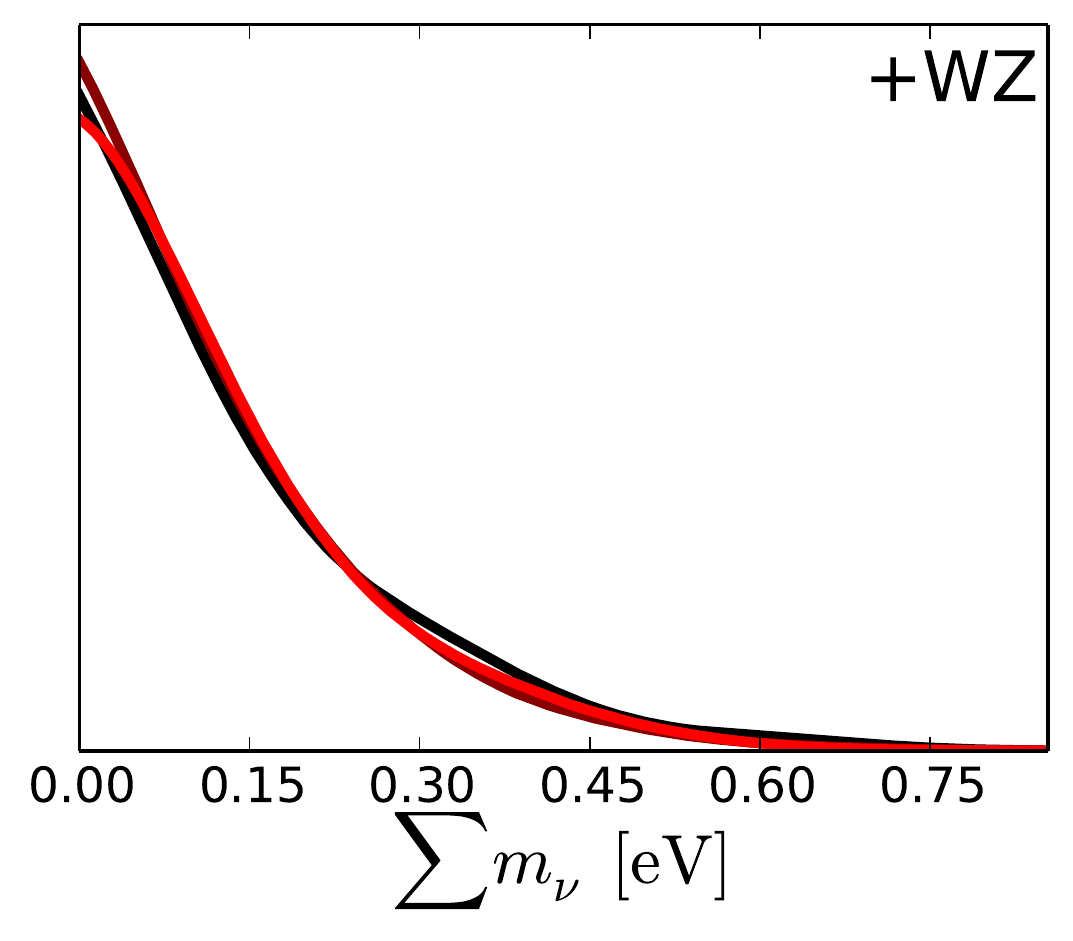}
\includegraphics[width=0.24\textwidth]{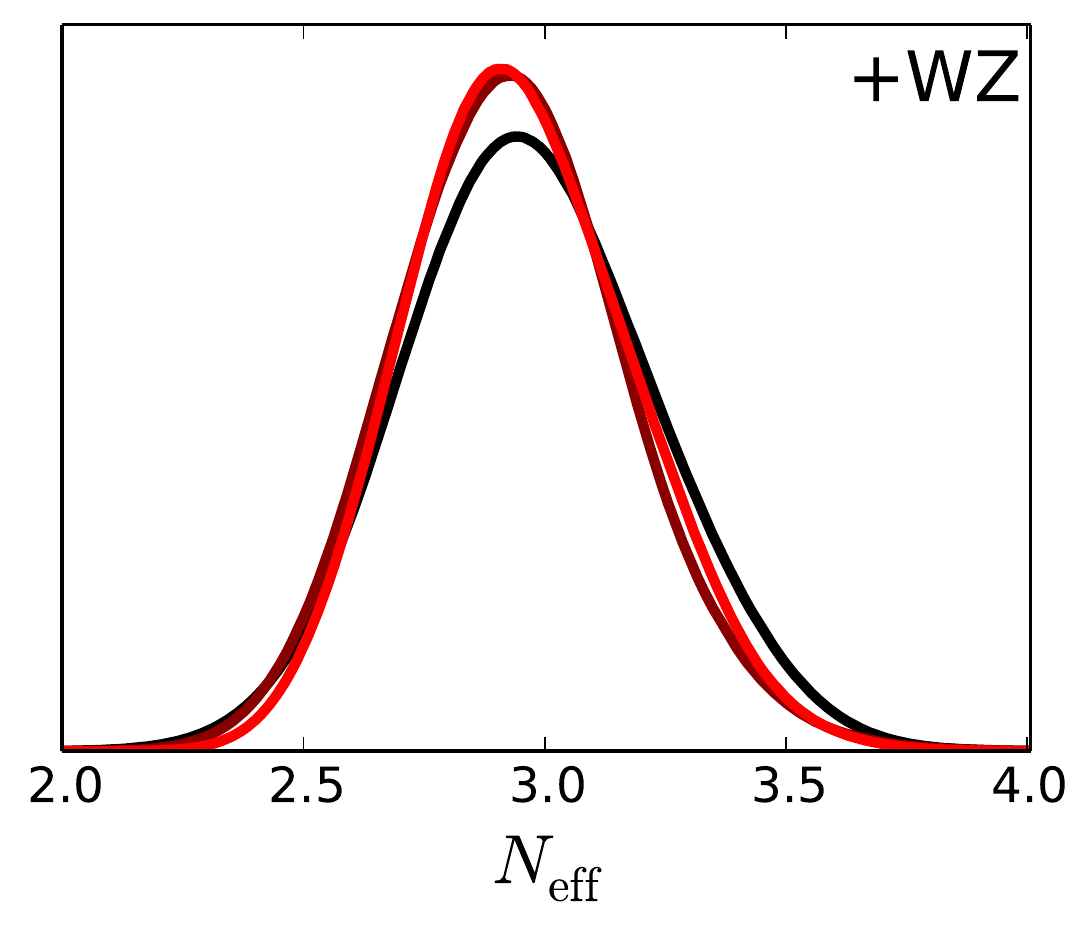}
\includegraphics[width=0.24\textwidth]{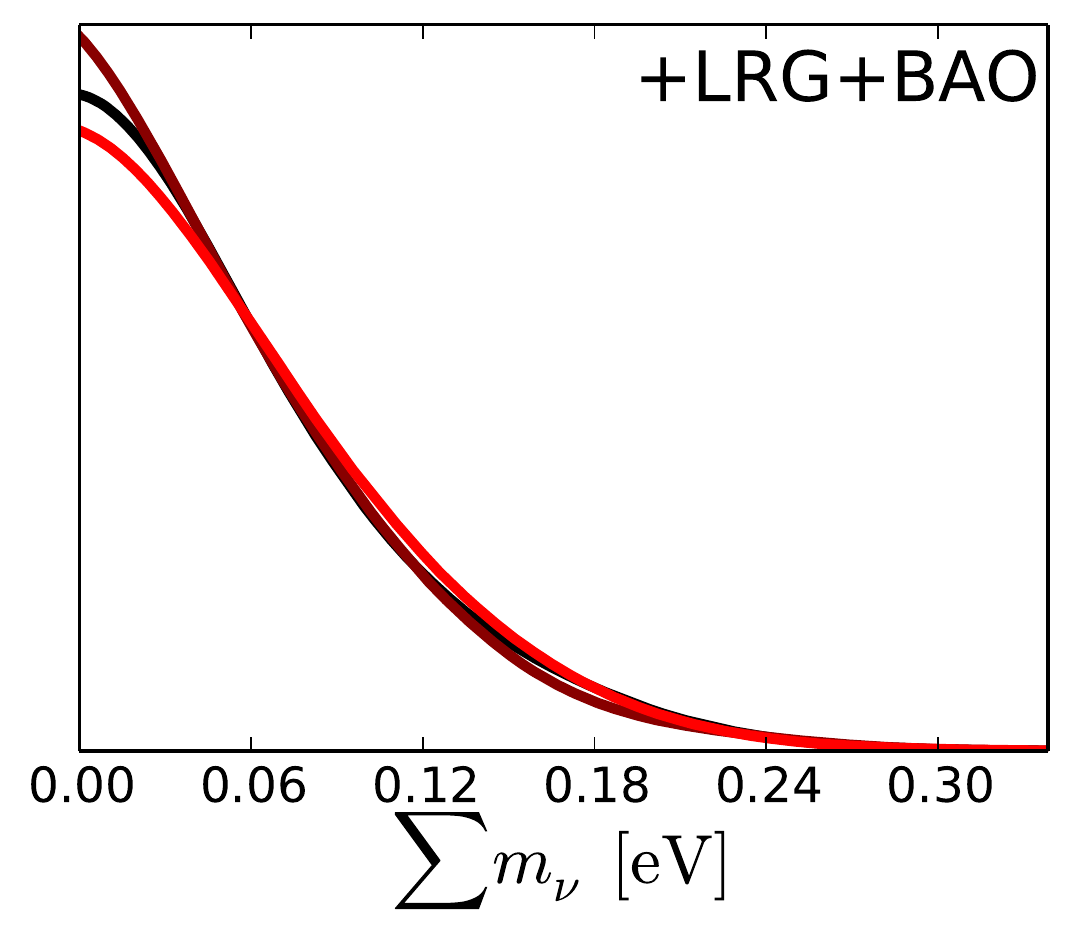}
\includegraphics[width=0.24\textwidth]{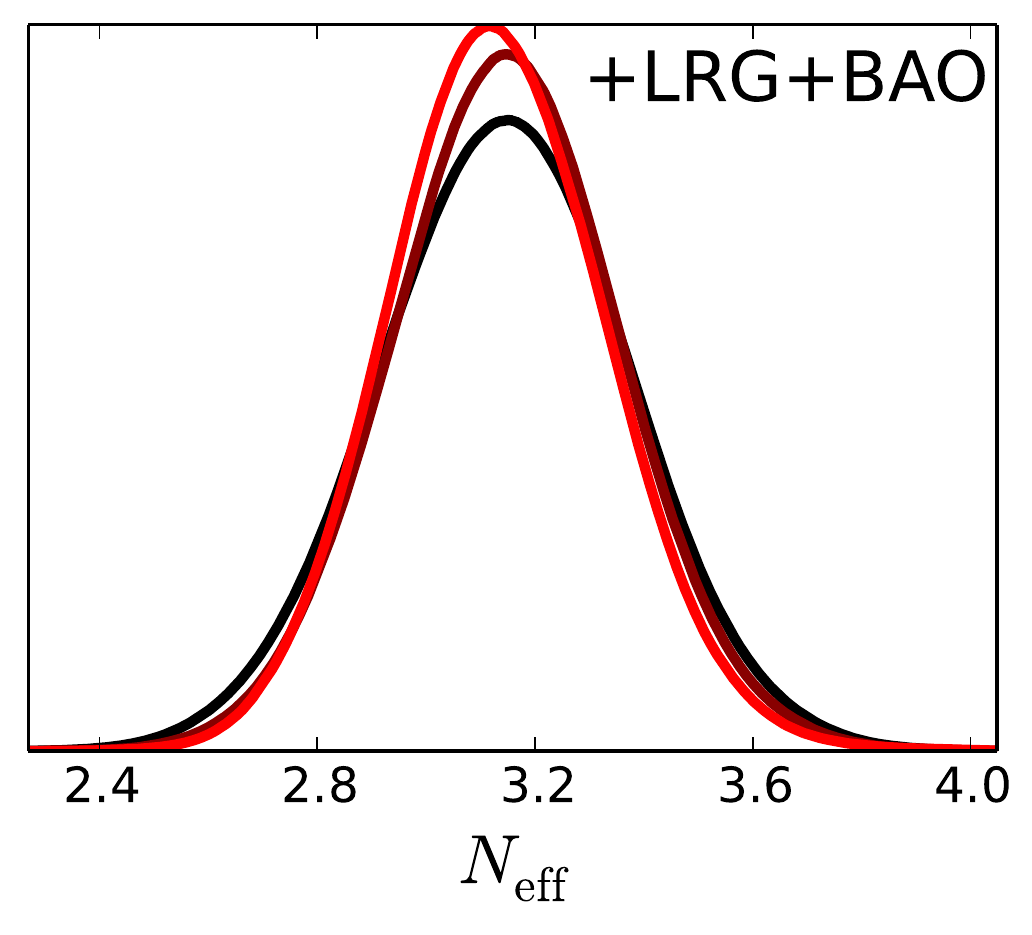}
\includegraphics[width=0.24\textwidth]{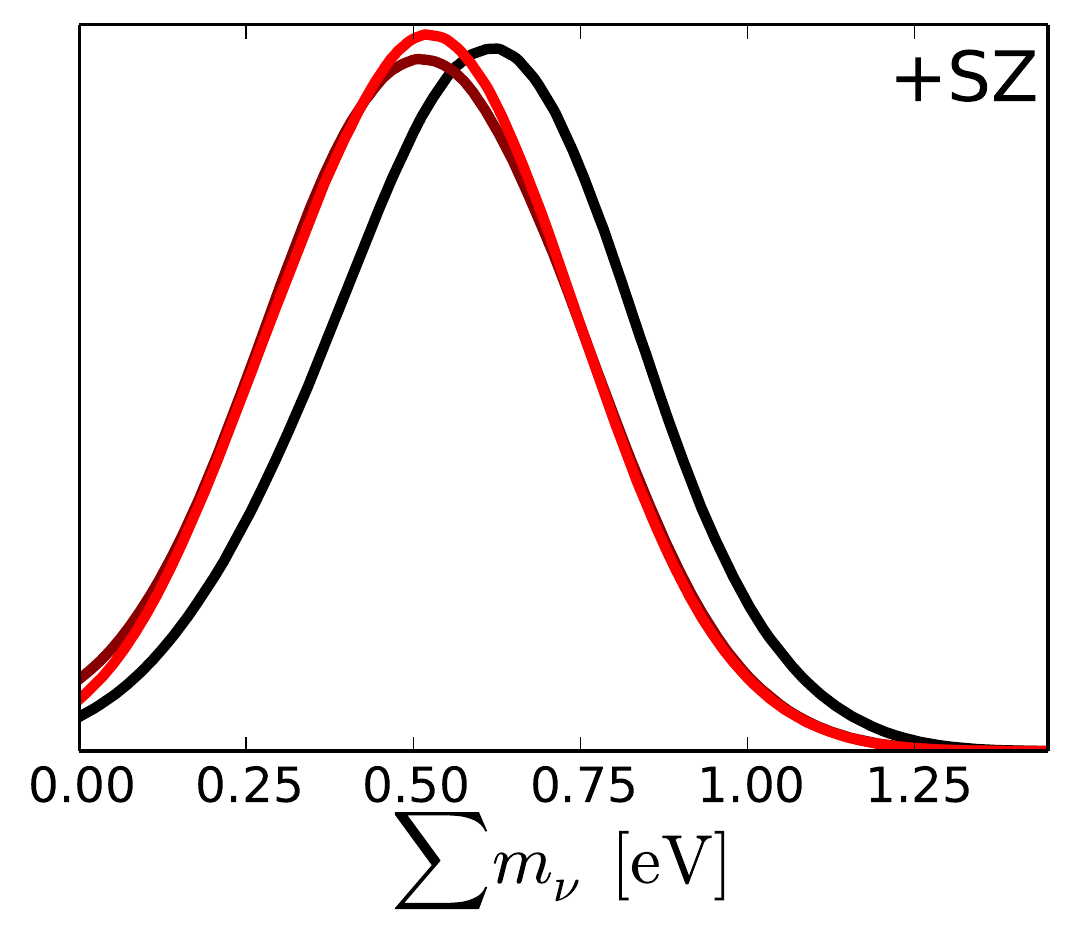}
\includegraphics[width=0.24\textwidth]{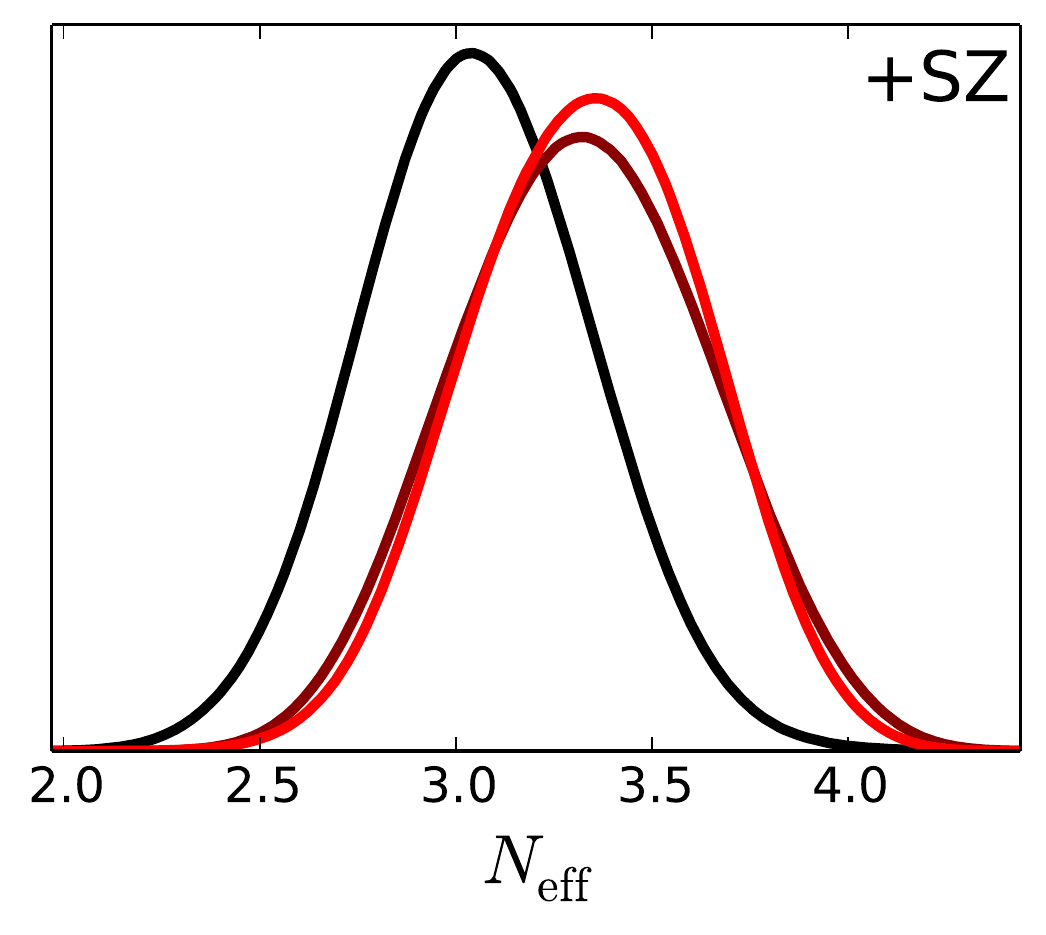}
\includegraphics[width=0.24\textwidth]{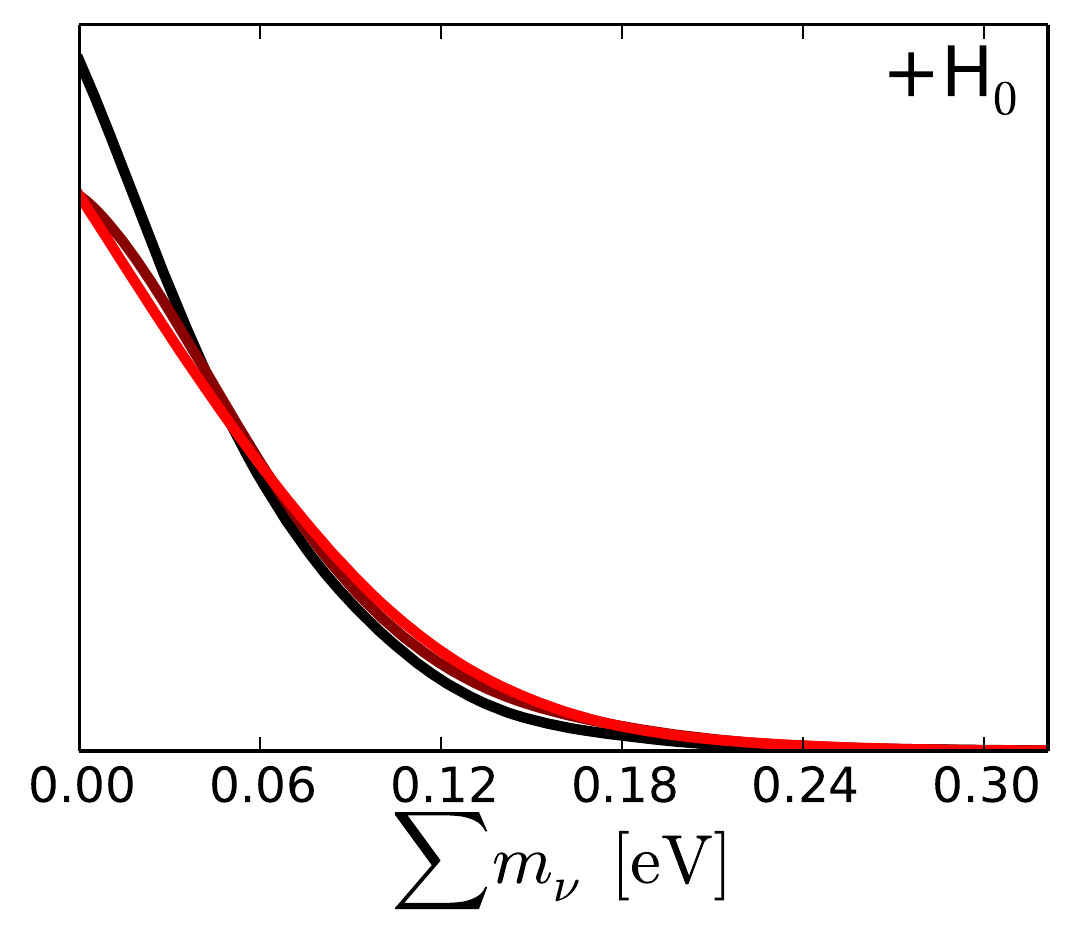}
\includegraphics[width=0.24\textwidth]{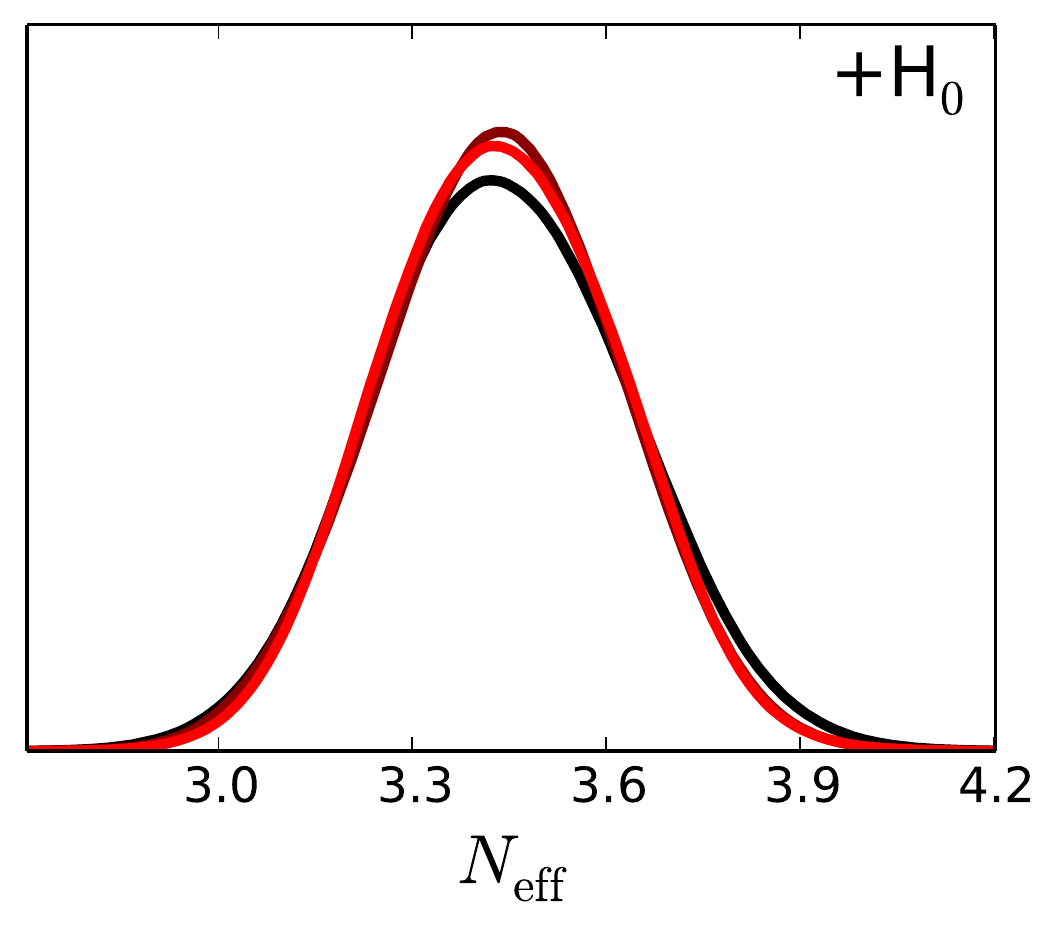}
\includegraphics[width=0.24\textwidth]{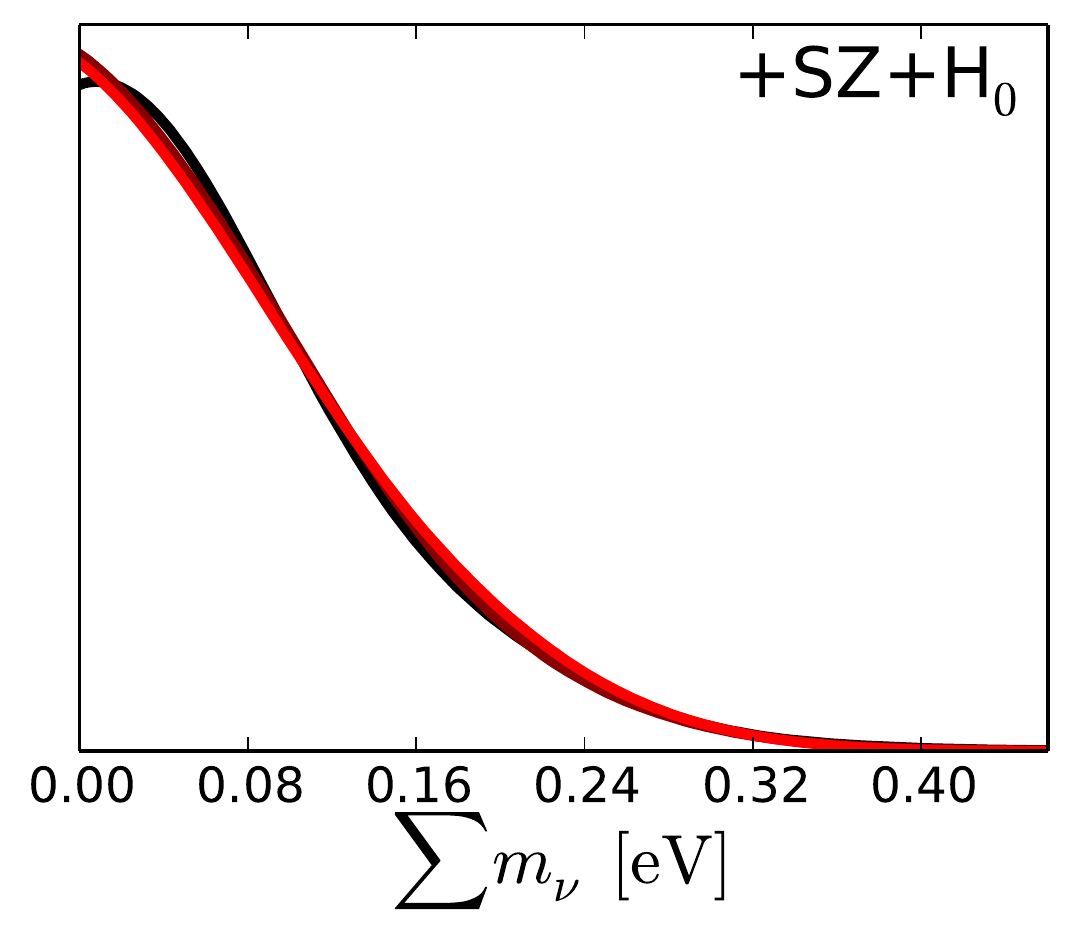}
\includegraphics[width=0.24\textwidth]{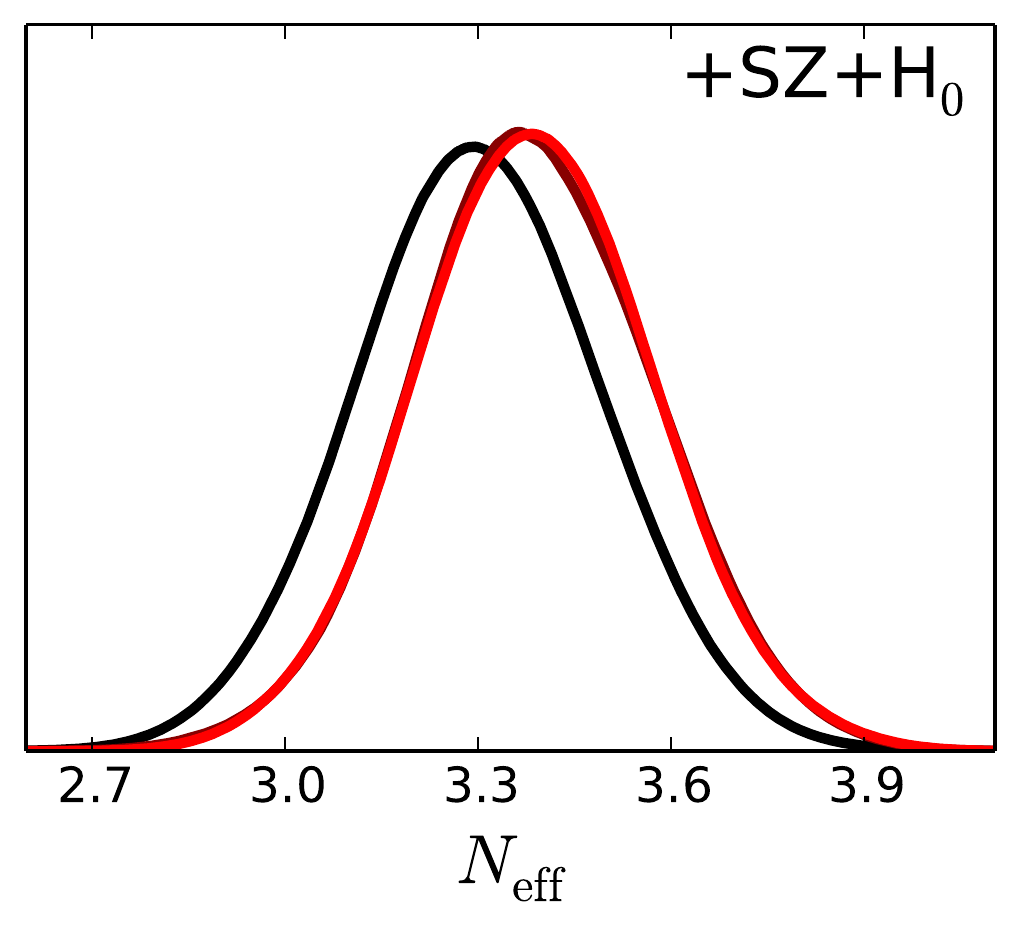}
\caption{\label{posterior_planck_wigglez_bao_fig}
One-dimensional posterior distributions for $\SigmaMnu$ and $N_\mathrm{eff}$. 
The colors of the contours from darkest to lightest indicate increasing number 
of knots in the PPS from 0 to 2 knots. Higher numbers 
of knots do not lead to significant changes. All cases shown are for 
models where the knot location in $k$ is a free parameter.
}
\end{figure}

\begin{table}
\renewcommand{\arraystretch}{1.5}
\setlength{\arraycolsep}{5pt}
\centering
\begin{tabular}{c c | c c}
  \hline
  \hline
  Parameter & Prior & Parameter & Prior \\
  \hline
  $\Omega_bh^2$ & $[0.020, 0.025]$ & $\SigmaMnu$ & $[0.001, 3.0]$ \\
  $\Omega_ch^2$ & $[0.10, 0.14]$ & $N_\mathrm{eff}$ & $[2.0, 5.0]$ \\
  $h$ & $[0.55, 0.80]$ & $\log_{10} k_\mathrm{knot}$ & $[-6, 1]$ \\
  $\tau$ & $[0.04, 0.12]$ & $\log (10^{10} \Delta_\mathrm{knot}^2)$ & $[-2, 4]$ \\
  \hline
  \hline
\end{tabular}
\caption{Ranges for uniform priors for cosmological parameters. 
}
\label{priors_table}
\end{table}

\begin{table}
\renewcommand{\arraystretch}{1.5}
\setlength{\arraycolsep}{5pt}
\centering
\begin{tabular}{c | c c}
  \hline
  \hline
  $\log $ (\emph{Posterior Odds}) &  Jeffreys Scale & Cosmology Scale \\
\hline
0.0 to 1.0 &  \multicolumn{2}{c}{Not worth more than a bare mention}  \\
1.0 to 2.5 &  Substantial & Weak \\
2.5 to 5.0 &  Strong & Significant \\
$> 5$      &  Decisive & Strong \\
  \hline
  \hline
\end{tabular}
\caption{Rough guideline for Bayesian evidence interpretation with the Jeffreys 
scale \cite{Jeffreys:1961} and the more conservative ``cosmology scale'' from 
Ref. \cite{hobson2010bayesian}.
}
\label{jeffreys_table}
\end{table}

\begin{figure}
\centering
\includegraphics[width=7.3cm]{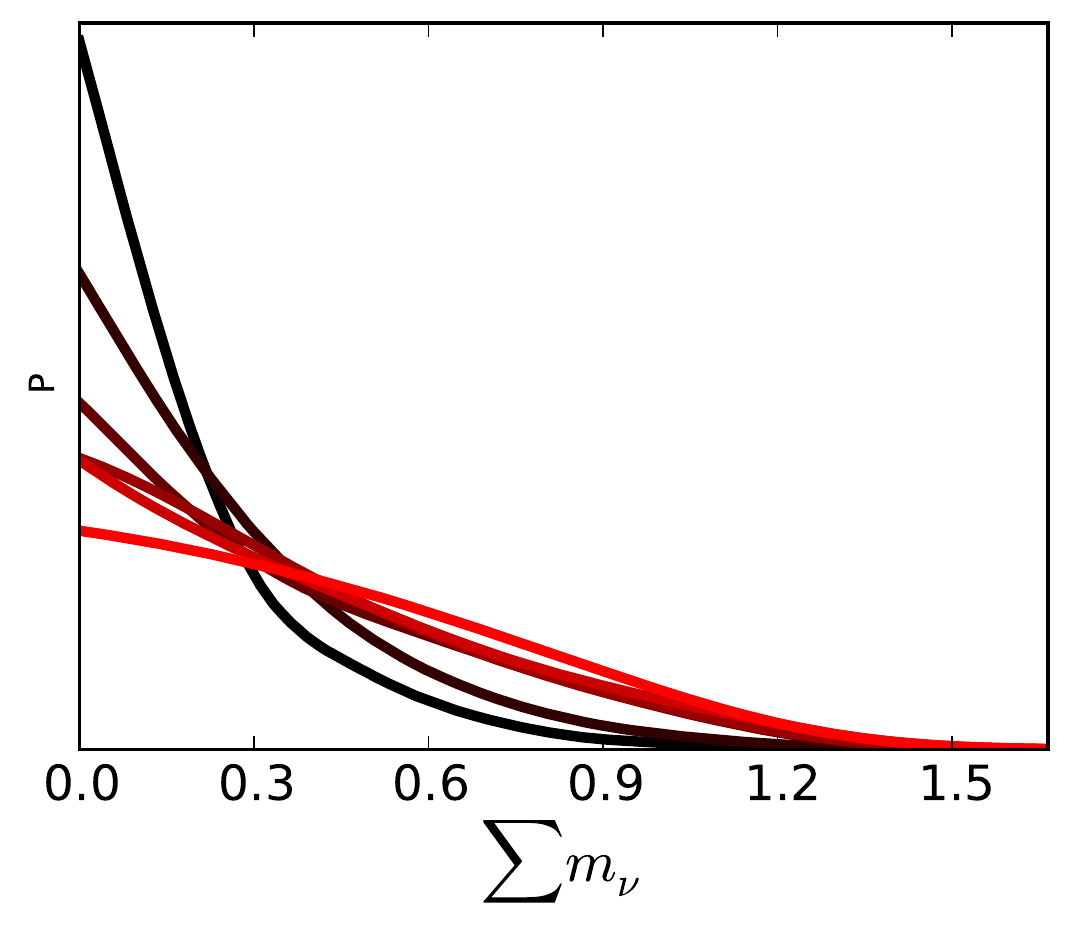}
\includegraphics[width=7.3cm]{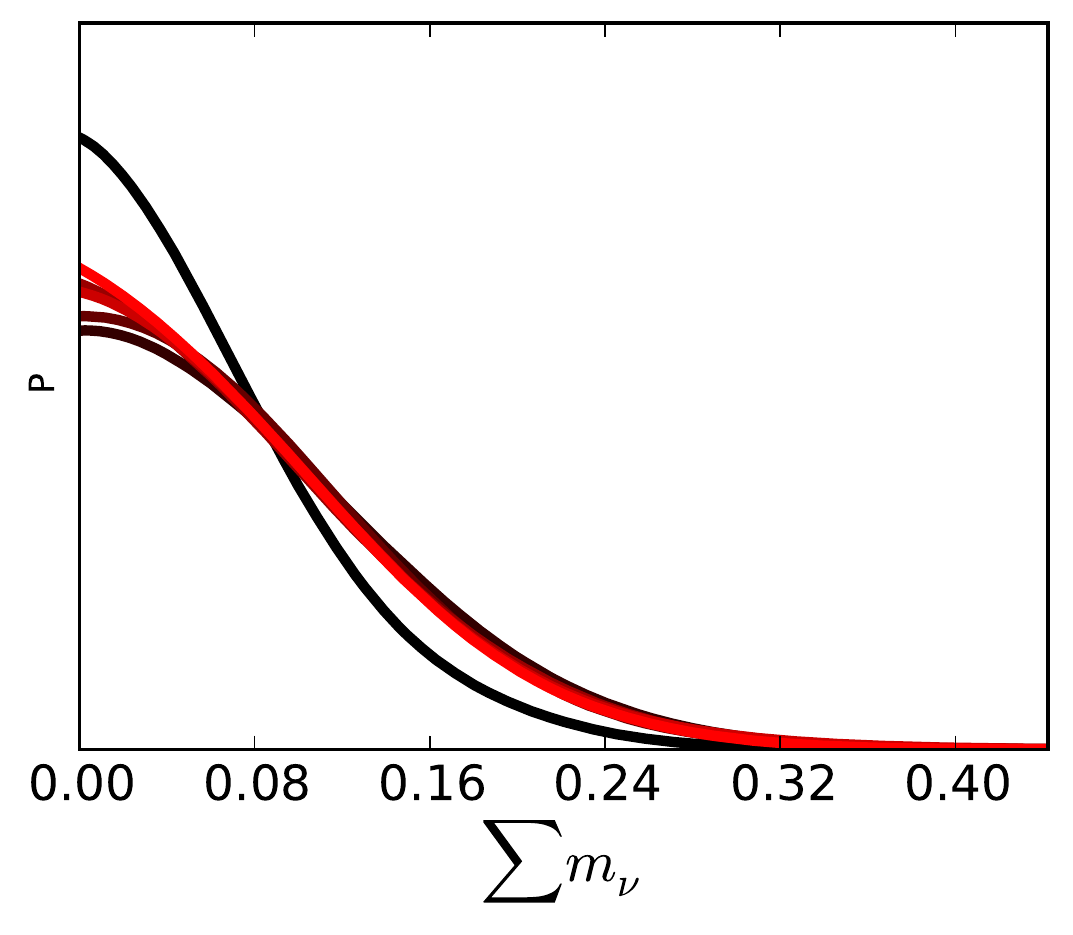}
\caption{\label{posterior_sum_fixed_fig} One-dimensional posterior 
distributions for $\SigmaMnu$ for (\emph{left}) Planck15 only and 
(\emph{right}) Planck15+BAO for the case of fixed knots. The color of line from 
dark to light represents increasing numbers of knots from 0 to 10 in intervals 
of 2.
}
\end{figure}

\begin{table}
\renewcommand{\arraystretch}{1.5}
\setlength{\arraycolsep}{5pt}
\centering
\begin{tabular}{c | c c c c c c}
  \hline
  \hline
  Data & No knots & 1 knot & 2 knots & 3 knots & 4 knots & 5 knots \\
  \hline
  Planck15 & $0.63$ & $0.66$ & $0.71$ & $0.64$ & $0.71$ & $0.67$ \\
  Planck15+BAO & $0.18$ & $0.19$ & $0.19$ & $0.18$ & $0.20$ & $0.24$ \\
  Planck15+WZ & $0.44$ & $0.40$ & $0.40$ & $0.39$ & $0.43$ & $0.37$ \\
  Planck15+LRG & $0.40$ & $0.40$ & $0.43$ & $0.39$ & $0.39$ & $0.38$ \\
  Planck15+$\mathrm{H}_0$ & $0.13$ & $0.15$ & $0.15$ & $0.18$ & $0.15$ & $0.18$ \\
  Planck15+WZ+BAO & $0.18$ & $0.18$ & $0.18$ & $0.17$ & $0.17$ & $0.17$ \\
  Planck15+LRG+BAO & $0.17$ & $0.16$ & $0.17$ & $0.17$ & $0.16$ & $0.22$ \\
  P15+LRG+BAO+$\mathrm{H}_0$ & $0.14$ & $0.13$ & $0.13$ & $0.16$ & $0.17$ & $0.13$ \\
  Planck15+SZ+$\mathrm{H}_0$ & $0.23$ & $0.23$ & $0.23$ & $0.24$ & $0.23$ & $0.24$ \\
  \hline
  Planck15+SZ & $0.60^{+0.22}_{-0.23}$ & $0.52^{+0.22}_{-0.24}$ & $0.52^{+0.22}_{-0.23}$ & $0.53^{+0.21}_{-0.23}$ & $0.49^{+0.20}_{-0.23}$ & $0.49^{+0.20}_{-0.22}$ \\
  \hline
  \hline
\end{tabular}
\caption{$95\%$ CI upper limits on $\SigmaMnu$ in eV for various
  combinations of data and numbers of knots for the model cases where
  $\SigmaMnu$ is allowed to vary in addition to the standard
  cosmological parameters and knots with $k$ and amplitude freedom in position. 
  The $68\%$ CI constraints are also shown in the bottom row for the
  tension data set.  }
\label{summnu_upper_limit_table}
\end{table}

\begin{table}[]
\renewcommand{\arraystretch}{1.5}
\setlength{\arraycolsep}{5pt}
\centering
\begin{tabular}{c | c c c c c c}
  \hline
  \hline
  Data & No knots & 1 knot & 2 knots & 3 knots & 4 knots & 5 knots \\
  \hline
  Planck15 & $3.14^{+0.31}_{-0.31}$ & $3.20^{+0.29}_{-0.29}$ & $3.22^{+0.27}_{-0.30}$ & $3.19^{+0.27}_{-0.29}$ & $3.25^{+0.26}_{-0.28}$ & $3.22^{+0.26}_{-0.29}$ \\ 
  Planck15+BAO & $3.14^{+0.24}_{-0.24}$ & $3.15^{+0.21}_{-0.21}$ & $3.15^{+0.21}_{-0.23}$ & $3.14^{+0.21}_{-0.23}$ & $3.14^{+0.20}_{-0.20}$ & $3.14^{+0.19}_{-0.21}$ \\
  Planck15+WZ & $2.97^{+0.26}_{-0.27}$ & $2.93^{+0.23}_{-0.24}$ & $2.95^{+0.22}_{-0.25}$ & $2.91^{+0.22}_{-0.25}$ & $2.91^{+0.23}_{-0.27}$ & $2.90^{+0.22}_{-0.27}$ \\
  Planck15+LRG & $3.07^{+0.26}_{-0.28}$ & $3.05^{+0.24}_{-0.25}$ & $3.05^{+0.23}_{-0.26}$ & $3.04^{+0.24}_{-0.26}$ & $3.04^{+0.25}_{-0.26}$ & $3.05^{+0.23}_{-0.27}$ \\
  Planck15+$\mathrm{H}_0$ & $3.44^{+0.19}_{-0.20}$ & $3.44^{+0.19}_{-0.19}$ & $3.44^{+0.19}_{-0.19}$ & $3.44^{+0.17}_{-0.19}$ & $3.43^{+0.19}_{-0.18}$ & $3.43^{+0.19}_{-0.18}$ \\
  Planck15+WZ+BAO & $3.01^{+0.21}_{-0.20}$ & $3.02^{+0.19}_{-0.22}$ & $3.01^{+0.18}_{-0.19}$ & $3.00^{+0.19}_{-0.19}$ & $3.01^{+0.18}_{-0.18}$ & $3.01^{+0.19}_{-0.19}$ \\
  Planck15+LRG+BAO & $3.14^{+0.23}_{-0.23}$ & $3.15^{+0.20}_{-0.21}$ & $3.13^{+0.20}_{-0.20}$ & $3.15^{+0.21}_{-0.22}$ & $3.13^{+0.20}_{-0.19}$ & $3.14^{+0.21}_{-0.22}$ \\
  P15+LRG+BAO+$\mathrm{H}_0$ & $3.36^{+0.19}_{-0.19}$ & $3.36^{+0.17}_{-0.17}$ & $3.36^{+0.17}_{-0.17}$ & $3.36^{+0.17}_{-0.17}$ & $3.36^{+0.16}_{-0.18}$ & $3.36^{+0.17}_{-0.16}$ \\
  Planck15+SZ & $3.05^{+0.29}_{-0.29}$ & $3.33^{+0.33}_{-0.34}$ & $3.34^{+0.31}_{-0.31}$ & $3.33^{+0.31}_{-0.31}$ & $3.36^{+0.30}_{-0.30}$ & $3.39^{+0.29}_{-0.29}$ \\
  Planck15+SZ+$\mathrm{H}_0$ & $3.30^{+0.19}_{-0.18}$ & $3.39^{+0.19}_{-0.18}$ & $3.39^{+0.18}_{-0.18}$ & $3.38^{+0.19}_{-0.19}$ & $3.38^{+0.19}_{-0.18}$ & $3.39^{+0.19}_{-0.19}$ \\
  \hline
  \hline
\end{tabular}
\caption{$68\%$ CI constraints on $N_\mathrm{eff}$ for various
  combinations of data and numbers of knots for the model cases where
  $N_\mathrm{eff}$ is allowed to vary in addition to the standard
  cosmological parameters and knots with $k$ and amplitude freedom in
  position.  }
\label{neff_constraints_table}
\end{table}

\subsection{Data sets}

We use measurements of the CMB, the matter power spectrum, BAO, SZ
cluster counts, and Hubble constant $\mathrm{H}_0$, which we describe in detail below. Likelihood modules for each
of these data sets have been written for use with
\textsc{Cosmo++}.

\paragraph{Cosmic Microwave Background.---} For all the runs performed in
this analysis, we use the CMB measurements from the Planck 2015 data
release \cite{Adam:2015rua}.  Although better constraints are provided
by the latest intermediate results from Planck using low-$\ell$ HFI
polarization~\cite{Aghanim:2016yuo}, the data is not currently
public. We use the interface provided by \textsc{Cosmo++} to include
the Planck 2015 likelihood code and use the full Planck CMB
temperature power spectrum at multipoles $2 \le \ell \le 2500$ along
with the Planck low-$\ell$ polarization likelihood in the range $2 \le
\ell \le 29$.  This combination of data is generally referred to as
Planck TT+lowP.  In this paper, we will refer to this combination of
data as ``Planck15.''

\paragraph{Matter Power Spectrum.---} In addition to CMB data, we also
include power spectra measurements using two different data sets. The
first data set comes from the SDSS Data Release 7
\cite{Abazajian:2008wr}. We use the most recent measure of the power
spectrum of the reconstructed halo density field derived from a sample
of $110,576$ LRGs in Reid et al.~\cite{Reid:2009xm}. As in the original
analysis, we include modes up to an upper bound of $k_\mathrm{max} =
0.2 \, h \, \impc$, above which uncertainties in nonlinear corrections
to the matter power spectrum become significant. The lower
bound of $k_\mathrm{min} = 0.02 \, h \, \impc$ is a function of
the survey volume. 
We have rewritten the original Reid et al.\ likelihood code in C++ in
order to interface with \textsc{Cosmo++}. As in the original code, we
include the effects on the linear power spectrum due to BAO damping,
non-linear structure formation, and halo bias. We model each of these
corrections identically to the original code, with the one
difference being that we are using an updated version of
\textsc{Halofit} \cite{Takahashi:2012em, Bird:2011rb}. The fiducial
model files have also been updated to include the effects of this new
version of \textsc{Halofit}. We will use the shorthand ``LRG'' to
refer to this set of data.

The second data set used for the matter power spectrum comes
from the WiggleZ Dark Energy Survey, which provides a measurement of
the matter power spectrum at redshifts $z = 0.22$, $z = 0.41$, $z =
0.60$, and $z = 0.78$ \cite{Parkinson:2012vd}. In our analysis, we
include only modes that satisfy $0.02 \, h \, \impc < k < 0.2 \, h \,
\impc$, as is done for the LRG data. Again, we have rewritten the
WiggleZ likelihood code in C++ for use with \textsc{Cosmo++} and
compared the sampling results to those obtained with
\textsc{MontePython} \cite{Audren:2012wb} to verify its accuracy. In
this paper, we will use ``WZ'' to refer to this data set.

\paragraph{Baryon Acoustic Oscillations.---} We have  included BAO data
from the 6dFGS \cite{Beutler:2011hx}, from SDSS-MGS
\cite{Ross:2014qpa}, and from BOSS Data Release 11, from both the LOWZ
and CMASS samples \cite{Anderson:2013zyy}. We note that in cases where
we use both LRG and BAO data, we omit the SDSS-MGS BAO data set in
order to avoid double counting of information. For cases which do not
include LRG, all four data sets are used when incorporating BAO
measurements.  From here on, we will refer to this data set simply as
``BAO.''

\paragraph{SZ Cluster Counts.---} In addition, we include information from the 
detection of 189 clusters by \emph{Planck} via the Sunyaev-Zeldovich effect 
\cite{Ade:2013lmv}. Cosmological constraints were  deduced in the 
$\sigma_8-\Omega_m$ plane, which was found to be $\sigma_8(\Omega_M/0.27)^{0.3} 
= 0.764 \, \pm \, 0.025$ for the case where the hydrostatic bias $1-b$ was 
allowed to vary in the range $[0.7, 1.0]$. The inclusion of this data will be 
referred to as ``SZ.''

\paragraph{Hubble Constant.---} Finally, we include recent high-precision
measures of the local Hubble expansion from the Hubble Space Telescope
observations of Cepheid variables.  This data was used to measure the
local value of the Hubble Constant to 2.4\%, as $\mathrm{H}_0 = 73.02
\pm 1.79 \, \mathrm{km} \, \mathrm{s}^{-1} \, \mathrm{Mpc}^{-1}$
\cite{Riess:2016jrr}. This measurement will be referred to as
``$\mathrm{H}_0$.''  There are previous assessments of the local
Hubble expansion that prefer lower values of $\mathrm{H}_0$
\cite{Efstathiou:2013via}. We choose the Reiss et
al.~\cite{Riess:2016jrr} determination because it addresses much of
the issues raised in Efstathiou~\cite{Efstathiou:2013via}, and is the
highest precision measurement of $\mathrm{H}_0$ thus far.  This result
is also of interest since it indicates tension at low redshift,
potentially from the same or different new physics leading to the
tension in the cluster samples.  We include the measure of
$\mathrm{H}_0$ to test what it indicates for $\Lambda$CDM in
combination with Planck and cluster data.

\begin{figure}
\centering
\includegraphics[width=15cm]{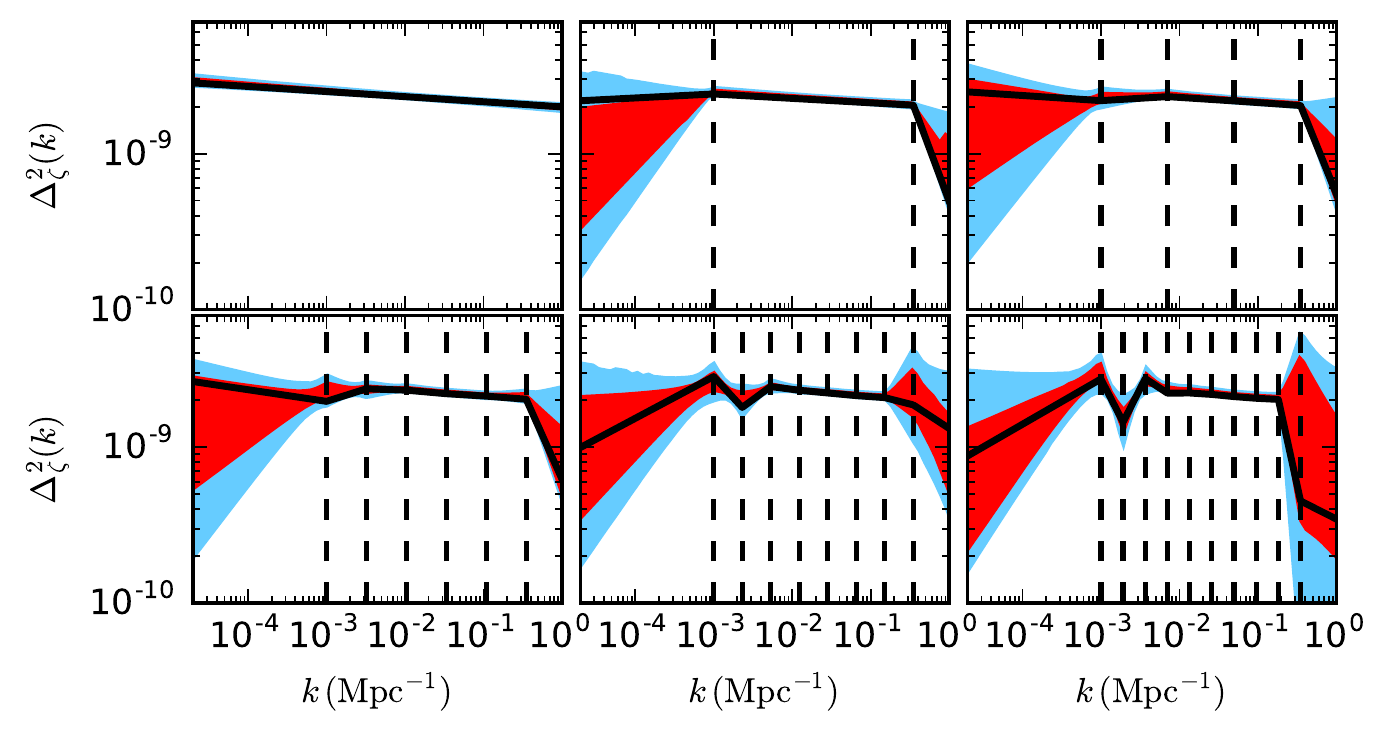}
\includegraphics[width=15cm]{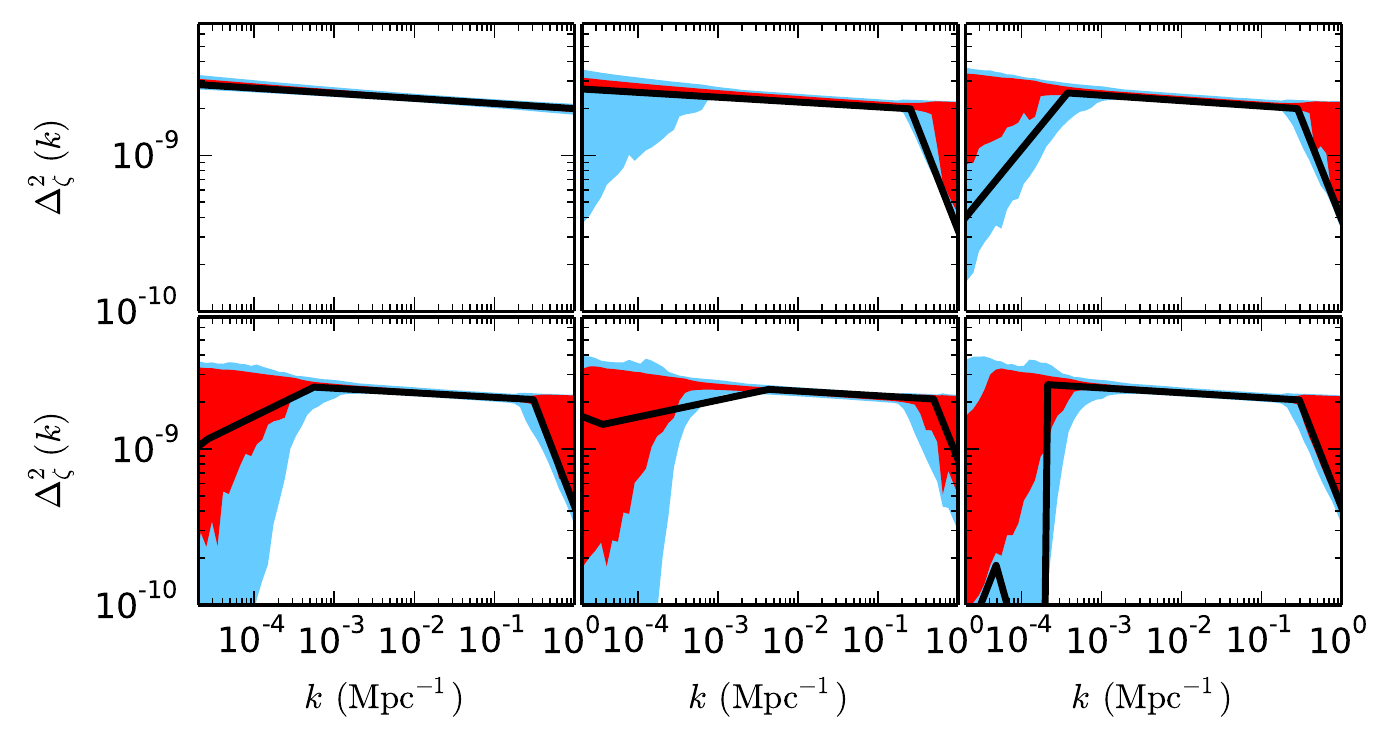}
\caption{\label{pps_planck_fixed_fig} Shown \emph{(top)} are the PPS
  estimated using Planck15 data for the case where the knot locations
  are fixed and only the amplitudes are varied, along with $\sum
  {m_\nu}$ to vary. The black line shows the best fit PPS, and the red
  and blue regions correspond to the $68\%$ and $95\%$ CI regions.  We
  see similar reconstructed power spectra when BAO is included
  \emph{(not shown)}. For comparison, we show \emph{(bottom)} the same
  PPS with between 0 to 5 knots, so the corresponding panels in the
  top and bottom figure have the same number of free parameters. }
\end{figure}

\section{Results and Discussion}

\begin{figure}
\centering
\includegraphics[width=15cm]{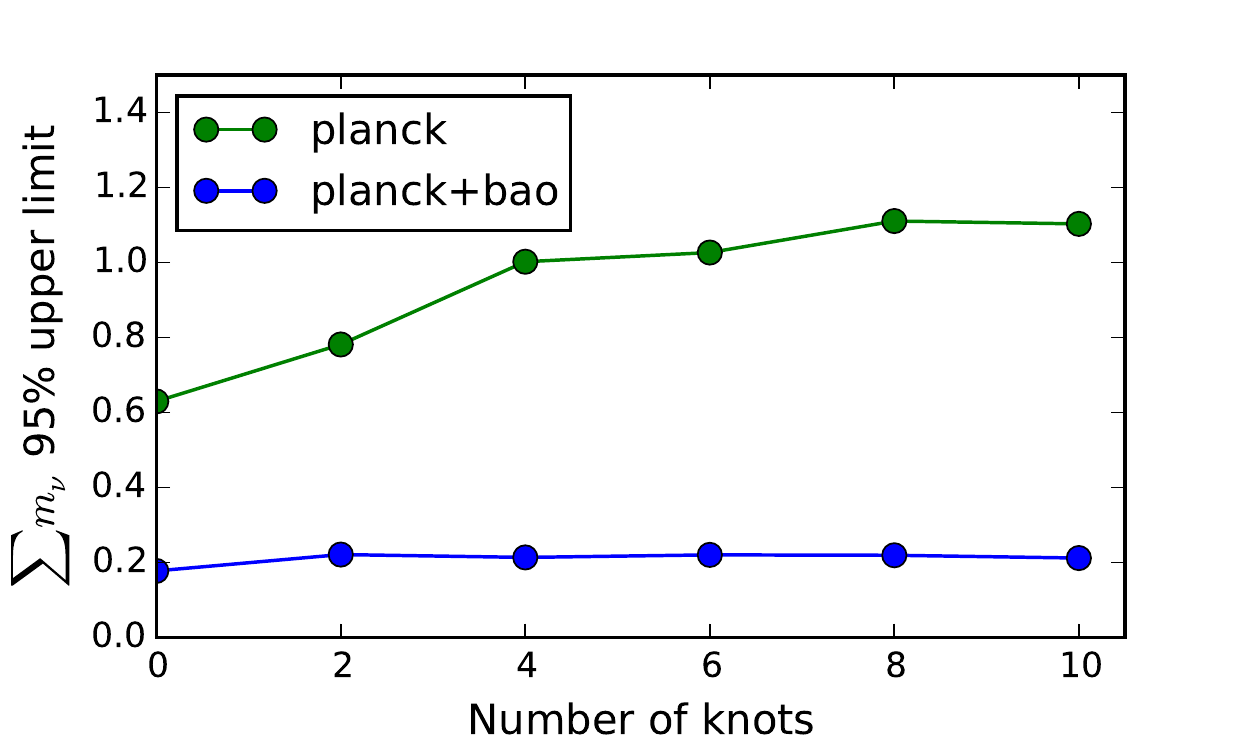}
\caption{\label{sum_upper_limits_fixed_fig} Shown are the $95\%$ CI upper 
limits on $\SigmaMnu$ for fixed knot locations when only Planck15 data is 
included and when Planck15+BAO are included.
}
\end{figure}

\paragraph{Neutrino masses and $\Neff$ with movable-knot PPS.---}
In Table~\ref{summnu_upper_limit_table}, we report the $95\%$ credible
interval (CI) upper limits on the sum of the neutrino masses for zero
to five knots for various combinations of data sets. The limits for
the case with zero knots are consistent with previous work examining
cosmological neutrino mass limits \cite{Cuesta:2015iho}. The last row
of Table~\ref{summnu_upper_limit_table} shows the $68\%$ CI
constraints on $\SigmaMnu$ when SZ is included along with Planck15.
This shows a preference for non-zero $\SigmaMnu$, indicating some
tension with the other data sets in this analysis. For each of these
cases, $\SigmaMnu$ is allowed to vary in addition to the other
standard cosmological parameters, along with the knot locations and
amplitudes. When the knot location is allowed to vary freely, the
constraints on $\SigmaMnu$ show no dependence on the number of knots
in our Monte Carlo analysis. Therefore, for reconstructions of this
form, constraints on $\SigmaMnu$ appear to be robust to changes in
the shape of the PPS.

Furthermore, our most stringent limit achieved using the combination
of data sets Planck15+LRG+BAO+$\mathrm{H}_0$ of $\SigmaMnu \lesssim 0.13 \,
\mathrm{eV}$ with one interior knot is comparable to the constraint
obtained with the $\Lambda$CDM spectrum (matching the same analysis in
Ref.~\cite{Giusarma:2016phn}).  This is beginning to approach the
minimum allowed value for the inverted neutrino mass hierarchy of
$\SigmaMnu \sim 0.1 \, \mathrm{eV}$ and suggests that future precision
measurements could distinguish between the two possibilities for the
neutrino mass hierarchy. Joint constraints on $\sigma_8$, $h$,
$\Neff$, and $\SigmaMnu$ shown in
Fig.~\ref{sigma8_v_summnu_0knots_fig}, illustrate how neutrino
constraints become tighter with the inclusion of additional data sets
and which data most significantly impact the constraints. We find that
inclusion of the BAO data provides the most significant improvement in
the limit on the sum of neutrino masses due to the complementarity of
BAO in constraining $\Omega_m$, which is highly degenerate with
$\SigmaMnu$.  The change in the posterior
distributions for varying numbers of knots and combinations of data
sets are shown in Fig.~\ref{sigma8_v_summnu_comparison_fig},
demonstrating again the fact that allowing for additional freedom in
the PPS does not significantly impact neutrino parameter
constraints. Knots beyond two are not shown as they do not change the
posterior distribution in a noticeable manner.

Table~\ref{neff_constraints_table} shows the derived constraints on
$N_\mathrm{eff}$ for zero to five knots for various combinations of
data sets.  In these cases, $N_\mathrm{eff}$ is allowed to vary along
with the other standard cosmological parameters, along with the PPS
knot locations and amplitudes.  As with $\SigmaMnu$, constraints on
$N_\mathrm{eff}$ also appear to be robust to a relaxation of the
assumption that the PPS adheres to a power-law shape. In all cases in
which $\mathrm{H}_0$ is included, the best fit value for $N_\mathrm{eff}$ is
observed to shift up, such that the standard value of $3.046$ lies
just outside the $95\%$ allowed region for the majority of
cases. Limits on $N_\mathrm{eff}$ are shown in
Fig.~\ref{neff_lims_fig}. The dependence of the one-dimensional
posterior distributions for $\SigmaMnu$ and $N_\mathrm{eff}$ are shown
in Fig.~\ref{posterior_planck_wigglez_bao_fig}. As mentioned, very
little change is present when going from 0 to 2 knots. Beyond 2
knots, there is no discernible change, and so only cases up to 2 knots
are shown.

\paragraph{Comparing fixed and movable-knot PPS.---}
To test how differences in methods for reconstructing the PPS
affect derived neutrino parameter constraints, we perform several runs
in which we fix the position of the knots in a similar manner to the
method used in Ref.~\cite{DiValentino:2016ikp}.
Fig.~\ref{posterior_sum_fixed_fig} shows how the posterior
distribution of $\SigmaMnu$ changes in the case where the position of
knots are fixed and only the amplitudes are allowed to vary. For these
cases, we follow a similar procedure to that described above for our
``knot-spline'' reconstruction, with the exception that the position
of the knots are fixed in $k$-space. We perform runs with $0$, $2$, $4$,
$6$, $8$, and $10$ knots, with locations indicated by the vertical
dotted lines in Fig.~\ref{pps_planck_fixed_fig}.

For the case where only Planck15 is included, the posterior
distribution of $\SigmaMnu$ varies considerably.  This contrasts with
our reconstruction in which the knot location is allowed to vary,
indicating that the neutrino parameter constraints are sensitive to
the prior on the allowed shape of the PPS. However, when information
from LSS is included, in this case in the form of BAO data, the degree
of change is significantly less. This is not surprising given the fact
that LSS is expected to be a much more sensitive probe of neutrino
mass than CMB measurements alone.  The $95\%$ upper limits on
$\sum{m_\nu}$ are shown in Fig.~\ref{sum_upper_limits_fixed_fig},
illustrating this difference. The PPS for our fixed-location knots are
shown in Fig.~\ref{pps_planck_fixed_fig}. These figures indicate more
of a preference for features in the PPS relative to the reconstruction
method in which the knot locations are allowed to vary freely.

\begin{sidewaysfigure}[ht]
\centering
\includegraphics[width=22.5cm]{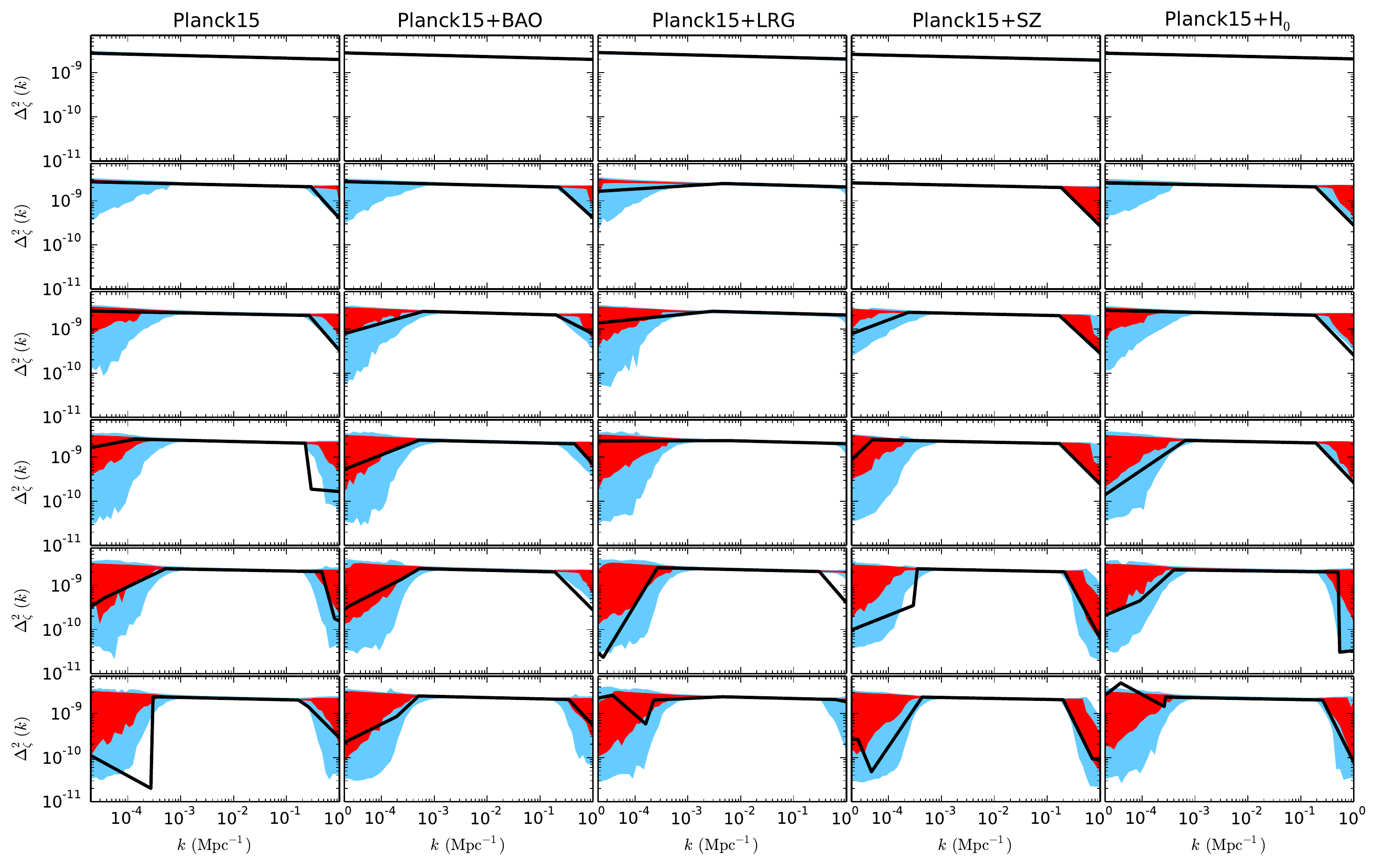}
\caption{\label{pps_large_fig}
    The reconstructed primordial power spectrum (PPS) without
  varying neutrino parameters. The columns correspond to different
  data sets used, shown at the top of each column. 
  The number of knots increases from $0$ to $5$ from
  top to bottom. The black solid lines show the best-fit PPS, the red
  lines are PPS in the $68\%$ CI, and the light blue lines are the PPS
  that fall in the $95\%$ CI.}
\end{sidewaysfigure}

In Fig.~\ref{pps_large_fig}, we show the reconstructed PPS for
cosmological models with different number of knots for Planck15 combined with 
various combinations of LRG, BAO, SZ, and $\mathrm{H}_0$ data sets. We find no 
significant features in the PPS using Planck15 with any combination of LRG, WZ, and BAO 
data. 
There is some apparent evidence for features when using the tension data sets 
SZ and $\mathrm{H}_0$ which will be discussed in more detail later in this paper. The
black lines represent the most likely power spectra. These all tend to
recover the standard power-law form for the PPS at small scales $k \,
\gsim \, 10^{-3} \, \impc$. At larger scales $k \lesssim 10^{-3} \,
\impc$, the best fit power spectra for models with non-zero numbers of
knots tend to indicate a suppression of power at large scales due to
the well-known low $C_\ell$ at low $\ell$ in the CMB (see, \emph{e.g.}
Ref.~\cite{Aslanyan:2014mqa}). Note that when allowing the knot positions to
freely vary, they accumulate in the cosmic variance dominated region,
so that functions with a fixed amount of variability will prefer to
fit large scale features preferably than small scale ones.  Since we do not have \emph{a priori} knowledge
of the position of features in the PPS, we allow the knot location to
vary. Furthermore, relaxing the position of the features protects
against the look-elsewhere effect or multiple comparisons problem,
since the knot is free to move over the global range of $k$.

\begin{figure}[h!b!]
\centering
\includegraphics[width=16cm]{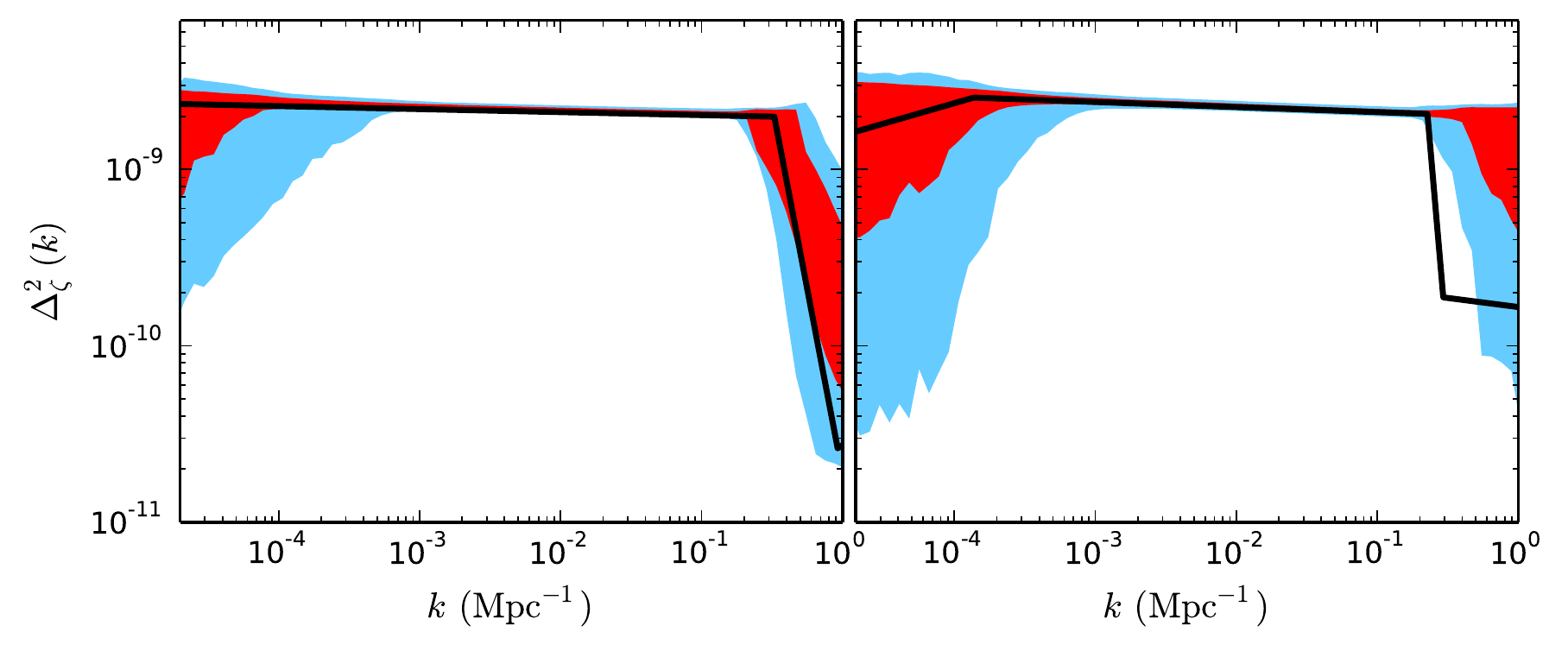}
\caption{\label{pps_sz_h0_best_fig}
Shown \emph{(left)} is the PPS for Planck15+SZ+$\mathrm{H}_0$ with 3 knots, 
which corresponds to our model with the highest evidence. For comparison, we 
also show \emph{(right)} the three knot case with only Planck15 data included.  
The black solid lines show the best-fit PPS, the red lines are PPS in the 
$68\%$ CI, and the light blue lines are the PPS that fall in the $95\%$ CI 
(note that it is possible for the PPS belonging to the sample with the 
maximum likelihood value to lie outside the red or blue regions).  }
\end{figure}

The confidence ranges for the power spectra in Figs. \ref{pps_planck_fixed_fig}, \ref{pps_sz_h0_best_fig} and \ref{pps_large_fig} are calculated as follows. 
For each $k$ value we construct a sample of all $P(k)$ values from our sample of power spectra, and from that sample we calculate the confidence intervals. 
The confidence intervals are constructed around the median of the sample, so that e.g. the $68\%$ confidence interval leaves out $16\%$ of the points of the 
sample on each side. This is the reason why the best fit lines sometimes lie outside the $68\%$ confidence range.

Table~\ref{bayes_factor_free_v_fixed_table} shows the $\Delta \ln(Z)$ values 
for the free and fixed-knot location PPS reconstructions described above with 
$\SigmaMnu$ free. In most cases, particularly for $\mathrm{DOF} \geq 
4$, Bayesian evidence strongly prefers the reconstruction in which the knot 
location is free over models with knots fixed at the positions shown in 
Fig.~\ref{pps_planck_fixed_fig}.
This indicates that a simple power-law fit 
over the approximate range $10^{-3} \, \mathrm{Mpc}^{-1} < k < 1 \, 
\mathrm{Mpc}^{-1}$ provides a significantly better fit than the hint of 
features seen in the PPS for fixed knot positions.

We should note that the big change in Bayesian evidences for the fixed knots case 
depends on the prior. The change would not be so dramatic if we had chosen a smaller prior
range for $\log(10^{10}\Delta^2_\mathrm{knot})$. Although we use the same prior for the moveable knots case, we do not see such large changes
in the evidence because the knots are able to move to parts of the parameter region
where there is high cosmic variance or noise, unless there are actual features to fit.
For the fixed knots case, however, there is only freedom to fit features at a priori fixed locations,
and if there are none then we see a significant drop in evidence.

\begin{table}
\renewcommand{\arraystretch}{1.5}
\setlength{\arraycolsep}{5pt}
\centering
\begin{tabular}{c | c c || c | c c}
  \hline
  \hline
  \multicolumn{3}{c}{Planck15} & \multicolumn{3}{c}{Planck15+BAO} \\
  \hline
  DOF & $\Delta \ln(Z)$ Free & $\Delta \ln(Z)$ Fixed & DOF & $\Delta \ln(Z)$ 
  Free & $\Delta \ln(Z)$ Fixed \\
  \hline
  $2$ & $1.26$ & $1.15$ & $2$ & $1.25$ & $1.24$ \\
  $4$ & $1.15$ & $-5.79$ & $4$ & $1.2$ & $-6.52$ \\
  $6$ & $1.48$ & $-13.03$ & $6$ & $1.42$ & $-13.5$ \\
  $8$ & $0.81$ & $-20.33$ & $8$ & $1.07$ & $-18.74$ \\
  $10$ & $0.47$ & $-21.59$ & $10$ & $1.21$ & $-22.47$ \\
  \hline
  \hline
\end{tabular}
\caption{Comparison of $\Delta \ln(Z)$ values relative to the $0$ knot case ($\Lambda$CDM) for free-form PPS models with knot position in $\log_{10} k$ either free or fixed.
For all cases, $\SigmaMnu$ is also free. The number of 
knots for each case is such that the number of additional degrees of freedom is 
equal in each row. For example, two additional degrees of freedom in the PPS 
corresponds to one knot in the free case (location and amplitude) and two knots 
in the fixed case (amplitude of each knot).
}
\label{bayes_factor_free_v_fixed_table}
\end{table}

\begin{table}
\renewcommand{\arraystretch}{1.5}
\setlength{\arraycolsep}{5pt}
\centering
\begin{tabular}{c | c || c | c}
  \hline
  \hline
  \multicolumn{2}{c}{Planck15+LRG+BAO} & \multicolumn{2}{c}{Planck15+LRG+BAO+$\mathrm{H}_0$} \\
  \hline
  Model & $\Delta \ln(Z)$ & Model & $\Delta \ln(Z)$ \\
  \hline
  $\Lambda$CDM & --- & $\Lambda$CDM & --- \\
  +1 knot & 0.99 & +1 knot & 1.60 \\
  +2 knots & 1.07 & +2 knots & 1.15 \\
  +3 knots & 0.75 & +3 knots & 1.40 \\
  +4 knots & 0.61 & +4 knots & 1.31 \\
  +5 knots & 0.59 & +5 knots & 0.78 \\
  --- & --- 
& +$N_\mathrm{eff}$+1 knot & 1.13 \\
  +$N_\mathrm{eff}$+2 knots & 0.10 
& +$N_\mathrm{eff}$+2 knots & 1.43 \\
  --- & --- 
& +$N_\mathrm{eff}$+3 knots & 0.74 \\
  --- & --- 
& +$N_\mathrm{eff}$+4 knots & 1.25 \\
  --- & --- 
& +$N_\mathrm{eff}$+5 knots & 0.89 \\
  \hline
  \hline
\end{tabular}
\caption{$\Delta \ln(Z)$ values relative to
  the six parameter $\Lambda$CDM model for various cosmological models for the combination of
  data sets Planck15+LRG+BAO and when $\mathrm{H}_0$ is added. Only models for
  which $\Delta \ln(Z)$ is positive relative to $\Lambda$CDM are
shown.}
\label{bayes_planck+lrg+bao+H0_table}
\end{table}

\paragraph{Model comparison.---}
The change in Bayesian evidence relative to the six-parameter $\Lambda$CDM model
is shown in Table~\ref{bayes_planck+lrg+bao+H0_table} for the data
sets Planck15+LRG+BAO(+$\mathrm{H}_0$). Importantly, there is no significant
preference for any model that includes $\SigmaMnu$, $N_\mathrm{eff}$,
or knots over a simple $\Lambda$CDM model.  The six-parameter $\Lambda$CDM model
is also preferred with Planck15 data alone.

However, with the inclusion of the recent measurement of $\mathrm{H}_0$, there 
is a significant preference for the combination of additional parameters
$N_\mathrm{eff}$ and knots. The cases with 2, 4, and 5 knots all satisfy 
$\Delta \ln(Z) > 2.5$, which can be interpreted as ``strong'' or
``significant'' odds against $\Lambda$CDM. The Bayes
factor for all models with positive $\Delta \ln(Z)$ for the Planck15+$H_0$ likelihood are shown in
Table~\ref{bayes_planck+H0_table}.

Similarly, when the SZ data set is included with Planck15, there is a
significant preference for the combination of a nonzero number of
knots along with a nonzero value for the sum of the neutrino masses
(see Table~\ref{summnu_upper_limit_table} for Planck15+SZ). When
$\mathrm{H}_0$ is also included, we see evidence for knots and
$N_\mathrm{eff}$. This comes largely from $N_\mathrm{eff}$ enhancing
the diffusion damping of the CMB anisotropies, which is recovered by a
larger $\mathrm{H}_0$ in order to preserve the position of the
acoustic peaks (see,
e.g. \cite{Hou:2011ec}). Table~\ref{bayes_planck+sz+H0_table} shows
the Bayes factor for both of these cases, Planck15+SZ and
Planck15+SZ+$\mathrm{H}_0$. There is a significant preference for
cosmological models with zero to five knots, with $\Delta \ln(Z) >
2.5$ for several model cases that contain additional non-standard
cosmological parameters. For the combination of data sets
Planck15+SZ+$\mathrm{H}_0$, a cosmological model with 3 knots is
heavily favored with $\Delta \ln(Z) > 5$. If we can assume that the
tension data set SZ is an accurate measurement, as well as the
measurement of $\mathrm{H}_0$, then these results represent strong
evidence that the PPS contains non-trivial features, not captured by a
power-law representation. In particular, for this case, we see a
preference for a very sharp cutoff at high $k$, as shown in
Fig.~\ref{pps_sz_h0_best_fig}. The cutoff in power at high $k$
provides the smaller amplitude in power indicated by the SZ
measurements. More conservatively, this is an indication that there
are systematic effects within the data that mimic PPS features, which
warrant further investigation. Importantly, smaller scale measures of
the PPS such as the Lyman-$\alpha$ forest would add more information
and likely disfavor strong departures from a power-law PPS
(e.g. \cite{Palanque-Delabrouille:2014jca}), but including the
inferred matter power spectrum from the Lyman-$\alpha$ forest is
beyond the scope of the current work.

\begin{table}
\renewcommand{\arraystretch}{1.5}
\setlength{\arraycolsep}{5pt}
\centering
\begin{tabular}{c | c || c | c}
  \hline
  \hline
  \multicolumn{2}{c}{Planck15} & \multicolumn{2}{c}{Planck15+$\mathrm{H}_0$} \\
  \hline
  Model & $\Delta \ln(Z)$ & Model & $\Delta \ln(Z)$ \\
  \hline
  $\Lambda$CDM & --- & $\Lambda$CDM & --- \\
  --- & --- 
& +$N_\mathrm{eff}$ & 0.49 \\
  +1 knot & 1.00 
& +1 knot & 1.56 \\
  +2 knots & 1.18 
& +2 knots & 2.02 \\
  +3 knots & 1.68 
& +3 knots & 1.72 \\
  +4 knots & 0.91
& +4 knots & 2.05 \\
  +5 knots & 0.60
& +5 knots & 1.92 \\
  +$N_\mathrm{eff}$+1 knot & 0.40 
& +$N_\mathrm{eff}$+1 knot & 2.27 \\
  +$N_\mathrm{eff}$+2 knots & 0.45 
& +$N_\mathrm{eff}$+2 knots & \emph{2.64} \\
  +$N_\mathrm{eff}$+3 knots & 0.13 
& +$N_\mathrm{eff}$+3 knots & 2.43 \\
  +$N_\mathrm{eff}$+4 knots & 0.50
& +$N_\mathrm{eff}$+4 knots & \emph{3.08} \\
  --- & ---
& +$N_\mathrm{eff}$+5 knots & \emph{2.69} \\
  \hline
  \hline
\end{tabular}
\caption{$\Delta \ln(Z)$ values relative to $\Lambda$CDM 
for various cosmological models for Planck15 and when $\mathrm{H}_0$ is added. Only models 
for which $\Delta \ln(Z)$ is positive relative to $\Lambda$CDM are shown.
}
\label{bayes_planck+H0_table}
\end{table}

\begin{table}
\renewcommand{\arraystretch}{1.5}
\setlength{\arraycolsep}{5pt}
\centering
\begin{tabular}{c | c || c | c}
  \hline
  \hline
  \multicolumn{2}{c}{Planck15+SZ} & \multicolumn{2}{c}{Planck15+SZ+$\mathrm{H}_0$} \\
  \hline
  Model & $\Delta \ln(Z)$ & Model & $\Delta \ln(Z)$ \\
  \hline
  $\Lambda$CDM & --- & $\Lambda$CDM & --- \\
  +$\SigmaMnu$ & 1.41 
& --- & --- \\
  +1 knot & \emph{2.65} 
& +1 knot & \emph{3.09} \\
  +2 knots & \emph{2.78} 
& +2 knots & \emph{3.48} \\
  +3 knots & \emph{3.06} 
& +3 knots & \textbf{5.03} \\
  +4 knots & \emph{4.37}
& +4 knots & \emph{4.74} \\
  +5 knots & \emph{4.33}
& +5 knots & \emph{2.94} \\
  +$\SigmaMnu$+1 knot & \emph{3.75} 
& --- & --- \\
  +$\SigmaMnu$+2 knots & \emph{3.49}
& --- & --- \\
  +$\SigmaMnu$+3 knots & \emph{3.33}
& +$\SigmaMnu$+3 knots & 1.60 \\
  +$\SigmaMnu$+4 knots & \emph{4.39}
& --- & --- \\
  +$\SigmaMnu$+5 knots & \emph{4.49}
& +$\SigmaMnu$+5 knots & 1.50 \\
  +$N_\mathrm{eff}$+1 knot & 2.05
& +$N_\mathrm{eff}$+1 knot & \emph{3.46} \\
  +$N_\mathrm{eff}$+2 knots & 2.28 
& +$N_\mathrm{eff}$+2 knots & \emph{3.36} \\
  +$N_\mathrm{eff}$+3 knots & 2.00 
& +$N_\mathrm{eff}$+3 knots & \emph{3.09} \\
  +$N_\mathrm{eff}$+4 knots & 1.76
& +$N_\mathrm{eff}$+4 knots & \emph{3.21} \\
  +$N_\mathrm{eff}$+5 knots & \emph{3.69}
& +$N_\mathrm{eff}$+5 knots & \emph{3.25} \\
  +$N_\mathrm{eff}$+$\SigmaMnu$+1 knot & 2.49 
& +$N_\mathrm{eff}$+$\SigmaMnu$+1 knot & 1.77 \\
  +$N_\mathrm{eff}$+$\SigmaMnu$+2 knots & \emph{2.60}
& +$N_\mathrm{eff}$+$\SigmaMnu$+2 knots & 2.17 \\
  +$N_\mathrm{eff}$+$\SigmaMnu$+3 knots & 2.46 
& +$N_\mathrm{eff}$+$\SigmaMnu$+3 knots & \emph{3.59} \\
  +$N_\mathrm{eff}$+$\SigmaMnu$+4 knots & \emph{2.81}
& +$N_\mathrm{eff}$+$\SigmaMnu$+4 knots & \emph{3.41} \\
  +$N_\mathrm{eff}$+$\SigmaMnu$+5 knots & \emph{3.71}
& +$N_\mathrm{eff}$+$\SigmaMnu$+5 knots & \emph{3.39} \\
  \hline
  \hline
\end{tabular}
\caption{$\Delta \ln(Z)$ values relative to
  6-parameter $\Lambda$CDM for various cosmological models for the
  combination of data sets Planck15+SZ and when $\mathrm{H}_0$ is added. Only
  models for which $\Delta \ln(Z) > 1.0$ relative to $\Lambda$CDM are
  shown. Significantly, there is no preference for neither extra
  relativistic degrees of freedom in $\Neff$ nor non-zero
  $\SigmaMnu$. }
\label{bayes_planck+sz+H0_table}
\end{table}

\section{Conclusion}

Using data from a broad set of the most recent cosmological
observations, including CMB, BAO, power spectrum, cluster counts, and
Hubble constant measurements, we have examined the dependence on the
assumed form of the PPS in measures of neutrino parameters $\SigmaMnu$
and $N_\mathrm{eff}$. To do this, we
applied the ``knot-spline'' method for reconstructing the PPS,
following \cite{Vazquez:2012ux,Aslanyan:2014mqa,Abazajian:2014tqa},
and allowing the knot positions to vary in location as well as
amplitude.  We found that for this method of reconstruction,
measures of $N_\mathrm{eff}$ and $\SigmaMnu$ do not appear to
depend strongly on assumptions about the PPS. However, when the knot
location is fixed, with CMB data alone, we observe a strong dependency
between $\SigmaMnu$ and the prior on the PPS. While including
information from LSS mitigates much of this dependency, this work
underscores the importance of quantifying the dependence of parameter
constraints on model assumptions and demonstrates the sensitivity of
neutrino parameter constraints on PPS priors and choice of data
sets. 

For combinations of data which include Planck15, BAO, LRG, and WZ, we
see no evidence for features in the PPS or a non-zero number of
knots. In addition, there is no preference for significant non-zero
neutrino mass or a value for $N_\mathrm{eff}$ outside of the standard
value expected in $\Lambda$CDM. Significantly, when we include recent
high-precision measurements of the low-redshift Hubble constant, we
find no significant evidence for extra relativistic energy density
$\Neff$. However, we do see relatively significant evidence for a
non-zero number of knots in concert with a value for $N_\mathrm{eff}$
that diverges from the standard value. When including the tension data
from SZ cluster counts, we see weak evidence for non-zero neutrino
mass and more significant evidence for knots or knots with
$N_\mathrm{eff}$ and $\SigmaMnu$.  However, when both $\mathrm{H}_0$
and SZ measurements are included, the preference for $\SigmaMnu$
vanishes, and only models which allow for both $N_\mathrm{eff}$ and
knots to vary are favored, with a model containing only 3 knots in
addition to the standard $\Lambda$CDM cosmological parameters being
preferred the most strongly over $\Lambda$CDM.  The radical difference
in the Bayesian evidences obtained for extensions to the 6-parameter
$\Lambda$CDM model with combinations of these data most conservatively
points to some unmodeled systematic effect, rather than a coherent
body of evidence in favor of non-standard cosmology. The tension data
could also indicate non-standard background expansion histories
\cite{DiValentino:2016hlg}.

As a combination of low, medium and high-redshift probes are
complementarily combined to constrain expansion history, cosmological
matter, dark energy, neutrino densities, and the primordial power
spectrum, robust methods of indications of new model features and
measures of new physics should be employed. Our work finds that
relaxing \emph{a priori} assumptions of the scales of features in the
primordial power spectrum does not significantly alleviate constraints
on neutrino mass and relativistic energy density. Significantly, we
find the tension in cosmological data from representative cluster data
sets do not significantly indicate a non-zero measure of massive
neutrinos. Also, we find the tension from Planck 2015 CMB and recent
high-precision $\mathrm{H}_0$ measures give no preference for a non-standard
$\Neff$. As cosmology enters an increasingly high-precision era with
multiple epoch and physical scale probes, robust statistical methods
and model tests will continue to be needed in order to make claims for
the discovery of new physics.

\acknowledgments

We thank Asantha Cooray, Elena Giusarma, and Manoj Kaplinghat for
useful discussions.  We acknowledge the use of the New Zealand
eScience Infrastructure (NeSI) high-performance computing facilities,
which are funded jointly by NeSI's collaborator institutions and
through the Ministry of Business, Innovation \& Employment's Research
Infrastructure programme [{\url{http://www.nesi.org.nz}}]. NC and KNA
are partially supported by NSF CAREER Grant No. PHY-1159224 and NSF
Grant No. PHY-1316792.  LCP is supported by the DOE DE-SC0011114
grant. KNA acknowledges hospitality and support by the Mainz Institute
for Theoretical Physics (MITP) program on ``Exploring the Energy
Ladder of the Universe'' where a portion of this work was completed.

\clearpage

\bibliographystyle{JHEP}
\bibliography{references}

\end{document}